\documentclass[journal]{IEEEtran}

\ifCLASSINFOpdf
 
\else
 
\fi

\usepackage{graphicx}
\usepackage{subcaption}
\usepackage{amsmath,amssymb}
\usepackage{booktabs,longtable,tabularx,array}
\usepackage{pdflscape}
\usepackage{enumitem}
\usepackage{microtype}
\newcolumntype{Y}{>{\raggedright\arraybackslash}X}
\newcolumntype{L}[1]{>{\raggedright\arraybackslash}p{#1}}
\begin{document}

\title{Graphene-based Hemispherical Transmitarray Antenna for Wide-Angle Beam Steering and Ultrafast Moving Target Tracking}

\author{Somayeh~Komeylian,~\IEEEmembership{Member,~IEEE}
        and~Christopher~Paolini,~\IEEEmembership{Member,~IEEE}
\thanks{S. Komeylian is with the Department of Electrical and Computer Engineering, 
University of California San Diego, La Jolla, CA 92093 USA, and the Department of Electrical and Computer Engineering, 
San Diego State University, San Diego, CA 92182 USA (e-mail: skomeylian@ucsd.edu).}
\thanks{C. Paolini is with the Department of Electrical and Computer Engineering, 
San Diego State University, San Diego, CA 92182 USA 
(e-mail: paolini@engineering.sdsu.edu).}}
\markboth{}%
{Shell \makeLowercase{\textit{et al.}}: Bare Demo of IEEEtran.cls for IEEE Journals}
\maketitle
\begin{abstract}
This work expands the application of dynamically tunable graphene to implement a hemispherical transmitarray antenna tailored for wide-angle electronic beam steering and ultrafast moving-target tracking at 250 GHz. 

The proposed hemispherical transmitarray antenna features a dynamically reconfigurable graphene-based multilayer configuration consisting of gold radiating patches, biased graphene sectors, hBN dielectric layers, and a centrally positioned hornfeed.

The analytical framework of graphene-based reconfigurable metasurfaces including graphene surface-conductivity modeling, voltage-controlled surface impedance, transmission-coefficient, and conformal array-factor analysis has been established for hemispherical transmitarray antenna for the first time. Moreover, the analytical and practical mapping between desired beam directions and the corresponding graphene bias voltages have been developed for the transmitarray antenna in the first time.  

Our transmitarray antenna possesses exceptional 3D beam steering, providing wide-angle elevation scanning from $-87^\circ$ to $87^\circ$ and full $360^\circ$ azimuthal coverage. It also exhibits an antenna efficiency ranging from 68\% to 80\% with a 10\% degradation. 
\end{abstract}
\IEEEpeerreviewmaketitle
\section{Introduction}
\IEEEPARstart{B}{eam} steering is a critical feature of next-generation antenna systems for THz frequency regime, providing adaptive directional radiation for high-capacity wireless communications, radar, and remote sensing applications, table 1. Various advanced beam-steering antenna technologies, including phased-array antennas (PAAs), reflectarrays, near-field array antennas, and transmitarray/metalens antennas, have been extensively developed and investigated by numerous research groups, \cite{reese2019millimeter,
tiwari2023active,
li2019wideangle,
alonsodelpino2024transmit,
omam2025wideband,
tiwari2024wideband,
madannejad2025passive,
jebreil2025fully,
keshavarz2026ultracompact,
jia2025lensbased,
madannejad2025silicon,
he2025crosstype,
alonsodelpino2024transmit_duplicate,
komeylian2021deep,
komeylian2023implementation,
komeylian2025hightechrxiv}.
Despite their excellent beam-steering performance, phased array antennas (PAAs) face several fundamental limitations associated with their conventional beamforming architectures, including high hardware complexity, substantial power consumption, and limited scalability for ultra-fast wide-angle scanning applications. Each antenna element typically requires an individual active phase shifter, and in many implementations, additional amplifiers or attenuators, leading to a significant increase in their system complexity, cost, and power consumption. Furthermore, the obtainable scanning speed is constrained by the switching response of phase shifters and the processing latency of beamforming control electronics. Real-time beam steering in PAAs requires continuous calculation, synchronization, and adjustment of the phase and amplitude excitation coefficients for all array elements, further increasing computational and hardware demands, \cite{he2025crosstype}.

Conventional reflectarray antennas typically achieve beam steering through mechanical displacement of their feeds or multiple feed switching to modify the incident phase distribution, rather than through electronically reconfiguration of the phase response of individual unit cells, \cite{balanis2016antenna,reese2019millimeter}. Hence, mechanical feed displacements limit their scanning speeds to the millisecond-to-second range, \cite{reese2019millimeter,tiwari2023active}. Even with multiple feedings, switching is generally restricted to discrete beam directions rather than continuous scanning. 
Furthermore, placing the feed antenna in front of the reflecting aperture causes aperture blockage, which drastically degrades efficiency, gain, and illumination uniformity, \cite{balanis2016antenna,alonsodelpino2024transmit}. This limitation becomes more pronounced at wide scanning angles, where severe phase errors, feed aberrations, and illumination non-uniformity exacerbate scanning loss and degrade beam quality, \cite{he2025crosstype,tiwari2024wideband}. As a result, the resonant nature of conventional unit cells strictly limits the operating bandwidth of many reflectarrays. This restriction limits the frequency range over which reflectarrays can maintain a stable phase response and high radiation efficiency, \cite{sharifi2022optimization,fallahi2010thin}.

Near-field metasurface beam-steering antennas achieve two-dimensional beam reconfigurations by incorporating engineered metasurface layers above the radiating aperture, enabling spatial control of the radiated electromagnetic wavefront. However, the addition of multiple metasurface layers to enhance phase manipulation capability increases the antenna profile and fabrication complexity, thereby limiting low-profile integration, \cite{tiwari2024wideband}, \cite{he2025crosstype}. Furthermore, electronically reconfigurable implementations often require complex biasing networks, while their obtainable scanning speed is limited by their response time of the tuning elements and associated control circuitry, \cite{tiwari2023active,li2019wideangle}. Moreover, the dielectric and conductor losses introduced by multilayer architectures can degrade radiation efficiency, particularly at millimeter-wave and terahertz frequencies, where material losses become increasingly significant, \cite{alonsodelpino2024transmit,he2025crosstype}.

Metalens antennas provide low-profile, high-integration solutions for beam focusing and steering by engineering the transmission phase profile of metasurface elements, \cite{li2019wideangle,alonsodelpino2024transmit}. However, conventional metalens antennas typically rely on fixed phase profiles or mechanical feed displacement, which severely restricts real-time reconfigurability and limits scanning speeds, \cite{reese2019millimeter,tiwari2023active}. Furthermore, single-focus metalenses compensate only for a first-order approximation of the required wavefront phase. This introduces substantial nonlinear phase errors at large steering angles, causing severe scan loss, a reduction in the realized gain, and a degradation in the beam quality during wide-angle operations, \cite{he2025crosstype,tiwari2024wideband,jebreil2025fully}. 
\begin{table}[htbp]
\centering
\caption{Comparison of beam-steering mechanisms and scanning speeds.}
\label{tab:beam-steering-speeds}
\scriptsize
\setlength{\tabcolsep}{3pt}
\begin{tabular}{c l c}
\toprule
\textbf{Rank} & \textbf{Technology} & \textbf{Speed} \\
\midrule
1 & Graphene voltage tuning (our work) & ns--$\mu$s \\
2 & Frequency-dependent passive beamforming [7,11] & ns--$\mu$s \\
3 & RF switch beam selection (LAS, mALL) [2,5,8,9] & ns--$\mu$s \\
4 & Digital beamforming with phase shifters [2,10] & ns--$\mu$s \\
5 & Mechanical feed displacement [4,13] & ms--s \\
6 & Reconfigurable metasurface lens replacement [1,6] & Offline (min) \\
\bottomrule
\end{tabular}
\end{table}
To overcome the aforementioned constraints, this work introduces a graphene-based hemispherical transmitarray antenna in Fig.~\ref{fig:transmitarray_antenna}. The main motivations and core contributions of this research are outlined in the following.
\begin{table}[t]
\centering
\scriptsize
\renewcommand{\arraystretch}{0.85}
\setlength{\tabcolsep}{2pt}
\caption{Performance comparison and categorization of state-of-the-art lens and metasurface antenna beam-reconfiguration methodologies.}
\label{tab:beam-reconfiguration-methods}
\begin{tabularx}{\columnwidth}{
L{0.16\columnwidth}
L{0.07\columnwidth}
L{0.12\columnwidth}
Y
L{0.10\columnwidth}
L{0.10\columnwidth}
L{0.18\columnwidth}}
\toprule
\textbf{Beam Reconfiguration Method} &
\textbf{Refs.} &
\textbf{Frequency Range} &
\textbf{Beam Steering Mechanism} &
\textbf{Scanning Range} &
\textbf{Scanning Speed} &
\textbf{Main Limitations}\\
\midrule

Electronic Switching &
[3,5,8,10] &
5.75--40 GHz &
PIN diodes, varactors, RF switches, multi-feed selection, optimized phase coding &
$\pm30^\circ$--$\pm60^\circ$ &
ns--$\mu$s &
High hardware complexity, insertion loss, and power consumption \\

\addlinespace

Mechanical Beam Steering &
[13] &
450--650 GHz &
Piezoelectric actuators translating a silicon lens or sparse lens array &
$\pm25^\circ$ &
Slow &
Moving mechanical parts and limited long-term reliability \\

\addlinespace

Passive Frequency-Dependent &
[11] &
610--685 GHz &
Frequency-dependent GRIN/Fresnel-lens phase-response interposer &
$16^\circ$--$32^\circ$ FoV &
Fast &
Beam direction strictly coupled to operating frequency \\

\addlinespace

Passive Fixed/Discrete &
[6,9] &
2.1--75 GHz &
1-bit binary metasurfaces, fixed phase distributions, DFrFT networks &
$\pm30^\circ$--$\pm60^\circ$ &
Static/Fast &
No continuous real-time beam tracking \\

\addlinespace

Feed-Position Multi-Focus &
[12] &
24.25--25.8 GHz &
Multi-focus phase compensation using feed-position switching &
$\pm48^\circ$ &
ms--s &
Complex feed configuration and limited flexibility \\

\addlinespace

Voltage-Controlled Graphene Tuning &
This work &
200--300 GHz &
Mapping desired beam direction to graphene bias voltages &
$\pm87^\circ$ elevation and $360^\circ$ azimuth &
Ultrafast: ns--$\mu$s &
Not reported \\
\bottomrule
\end{tabularx}
\end{table}

\textbf{Ultrafast Graphene-based Beam Steering and Real-Time Target Tracking}:
Our proposed graphene-based transmitarray antenna is specifically tailored for millimeter-wave (mmW) and the THz regime, overcoming the high propagation loss, narrow tuning bandwidths, and hardware complexity of conventional beam-steering antennas, Table~\ref{tab:beam-reconfiguration-methods}. 
A major advantage of graphene relies in its electrically tunable surface conductivity, enabling real-time modulation of the unit-cell surface impedance and phase response without incorporating conventional semiconductor active devices into the terahertz electromagnetic structure.
By independently tuning the chemical potential of the isolated graphene sectors, our transmitarray antenna achieves programmable phase and amplitude control, which allows for rapid electronic beam reconfiguration without mechanical scanning or feed displacement, Table~\ref{tab:beam-steering-speeds}.
This agile beam adaptation is highly advantageous for emerging 6G communication systems, high-resolution radar, and remote sensing systems that require rapid responses to dynamic propagation environments. 
Moreover, the fast reconfiguration capability, potentially reaching the nanosecond-to-microsecond regime and primarily determined by the graphene biasing network rather than mechanical limitations, enables rapid adjustment of the aperture phase distribution and beam direction. 
Consequently, the proposed antenna can maintain continuous beam alignment and compensate for angular displacements of moving targets over a wide steering range, Figs.~\ref{fig:scanning angles} and \ref{fig:scanning angles-1}.

In this work, the proposed transmitarray antenna will be integrated with the AMD Xilinx Zynq UltraScale+ RFSoC ZCU670 platform available in our laboratory to enable real-time tracking of moving targets through dynamic beam steering. A practical and scalable bias-control implementation is achieved by organizing the 121 graphene elements into an $11\times11$ voltage-control matrix. The FPGA computes the required $11\times11$ voltage-control matrix (or retrieves it from a precomputed lookup table) for a specified beam-steering angle and distributes the corresponding bias signals to multiple high-speed multichannel digital-to-analog converters (DACs) via parallel serial peripheral interface (SPI) or low-voltage differential signaling (LVDS) interfaces. The proposed row-parallel control architecture enables full-aperture reconfiguration within $10-20\mu s$, corresponding to 50,000-100,000 beam updates per second. This scalable implementation avoids the excessive hardware overhead and complexity of deploying an individual DAC channel for every graphene element while maintaining the high-speed reconfigurability required for adaptive beam steering and real-time target tracking, Table~\ref{tab:beam-steering-speeds}.

\textbf{High Gain and Broadband Hemispherical Transmitarray Aantenna}:
From a structural perspective, the hemispherical curvature of our antenna establishes an inherent phase-matching interface with the spherical wavefront emitted by the central feedhorn, thereby mitigating aperture phase errors and enhancing radiation efficiency, Table~\ref{tab:graphene_antenna_comparison}. 
Furthermore, scaling the dimensions of the fractal concentric circular (FCC) elements, Fig.~\ref{fig:FCC patches}, from row to row excites multiple coupled resonant modes with closely spaced frequencies. This yields overlapping resonances that significantly expand the operational bandwidth, Figs.~\ref{fig:scaling_factor} and \ref{fig:S11}. 

\textbf{Wide Spherical Coverage and Wide-Angle Beam Steering for 6G Wireless Communications}:
Spherical coverage defines the solid-angle range over which an antenna effectively radiates or receives electromagnetic energy, \cite{balanis2016antenna}. 
As wireless networks transition toward the THz spectrum for 6G communications, propagation conditions become increasingly challenging due to severe free-space path loss and atmospheric absorption. 
Although antenna arrays are traditionally deployed to mitigate these losses through high directional gain, such high-gain configurations inherently restrict the radiation beamwidth, and thereby angular coverage. 
Consequently, the achievable spherical coverage is significantly influenced by the antenna topology, beam-steering capability, and the interference characteristics of the wireless communication channel,
\cite{madannejad2025passive,
madannejad2025silicon,
tzarouchis2018light,
alonsodelpino2024transmit}. 
To address these limitations, our proposed three-dimensional hemispherical transmitarray antenna provides a vastly superior spherical coverage compared to conventional planar configurations, \cite{madannejad2025passive, madannejad2025silicon, tzarouchis2018light, alonsodelpino2024transmit}. Its curved architecture enables wide-angle beam steering over an broad field of view while maintaining high directive gain, Figs.~\ref{fig:scanning angles} and \ref{fig:scanning angles-1}, thereby elevating the average signal-to-interference ratio (SIR). Consequently, our antenna offers enhanced robustness against rapid channel variations, positioning it as a compelling solution for future THz-band 6G wireless communication networks.

\textbf{An Electromagnetic Modeling Framework for a Voltage-Reconfigurable Graphene Hemispherical Transmitarray Antenna}:
To the best of the authors' knowledge, this work is the first to establish a unified electromagnetic analytical framework for a multilayer graphene-based hemispherical transmitarray antenna. The framework analytically derives formulations for the voltage-controlled surface impedance, equivalent lumped RLC circuit model, transmission coefficient of $S_{21}$, antenna efficiency, quasi-active spherical array factor, and the mapping between desired beam directions and the corresponding graphene bias voltages, thereby enabling the systematic design and analysis of ultrafast beam-steering systems for real-time moving-target tracking. 

Consequentially, the remainder of this work is organized into the eight following sections: Section II describes the design parameters and geometric configuration of the proposed hemispherical graphene-based metasurface transmitarray antenna. 
Section III develops equivalent impedance models, including voltage-controlled graphene conductivity, multilayer equivalent circuit, and anisotropic surface impedance formulation. 
Section IV is devoted to the analytical formulation for extracting the transmission coefficient of $S_{21}$ of the multilayer transmitarray unit cell.
Section V establishes the analytical and practical mapping between the desired beam direction and the corresponding bias-voltage distribution applied to the graphene sectors. 
In section VI, we have derived the conformal hemispherical array factor and the beamforming formulation. 
The functionality and performance of our proposed metasurface transmitarray antenna have been verified and validated in section VII. 
Section VIII evaluates the transmission efficiency and the overall antenna efficiency of the proposed hemispherical transmitarray antenna. Finally, Section IX concludes this work and summarizes the principal contributions.
\section{Geometric Configuration and Design Parameters of the metasurface Tansmitarray Antenna}
\begin{figure}[htbp]
    \centering    \includegraphics[width=1\columnwidth,height=5cm,keepaspectratio]{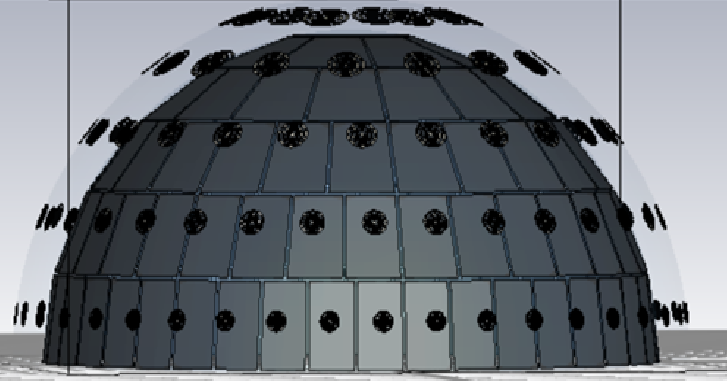}
    \caption{The hemispherical transmitarray antenna comprises 121 patch elements distributed over the hemispherical aperture, including one apex element and five concentric rings containing 8, 16, 24, 32, and 40 patches, respectively. A geometric scaling factor of 0.8 is applied to the patch dimensions from one ring to the next. Our stacked multilayer transmitarray antenna consists of $SiO_2$ substrate with a thickness of 0.5 mm, followed by the first hBN dielectric layer with a thickness of 0.5 mm, the graphene sectors, the second hBN dielectric layer with a thickness of 0.5 mm, and FCC patches made of gold (Au) on the top layer exposed to free space.}
    \label{fig:transmitarray_antenna}
\end{figure}

In this section, the design topology of the proposed metasurface transmitarray antenna is first presented. 
The stacked multilayer arrangement of our transmitarray antenna is introduced, and the functional role of each constituent layer, including the electrically tunable graphene layer, metallic FCC patch layer, hBN dielectric isolation layers, gate dielectric layer, and supporting substrate, is discussed in detail.
In the following, the key geometric design parameters of our proposed transmitarray antenna are systematically investigated, as they critically determine the electromagnetic behavior of individual unit cells by controlling their resonant characteristics, impedance response, and wave transmission properties.
Accordingly, this section investigates the geometric parameters of the gold FCC patches and graphene sectors, including their spatial distribution, dimensional scaling, and ring-to-ring dimensional variation of the unit cells across the hemispherical lens aperture.
It is important to note that, although the graphene sectors and metallic FCC patches are spatially aligned with coincident centers, they remain electrically isolated through an intermediate hBN dielectric layer. This insulating hBN layer prevents direct electrical contact between the two conductive materials, thereby avoiding unintended short circuits while preserving independent electromagnetic control of each unit cell, as illustrated in Fig.~\ref{fig:transmitarray_antenna}.
\subsubsection{Stacked Multilayer Configuration of the Transmitarray Antenna}
The multilayer stack configuration of the our transmitarray antenna, Fig. 1, consists of the following layers.

\textbf{FCC gold pathes}: The gold FCC patch frequency-selective surface (FSS) acts as the transmitting resonant layer and controls the amplitude and phase response of each unit cell. This layer, including gold FCC patches, radiates into the surrounding free-space region (outside air).

\begin{figure}[htbp]
    \centering
    \includegraphics[width=0.7\linewidth]{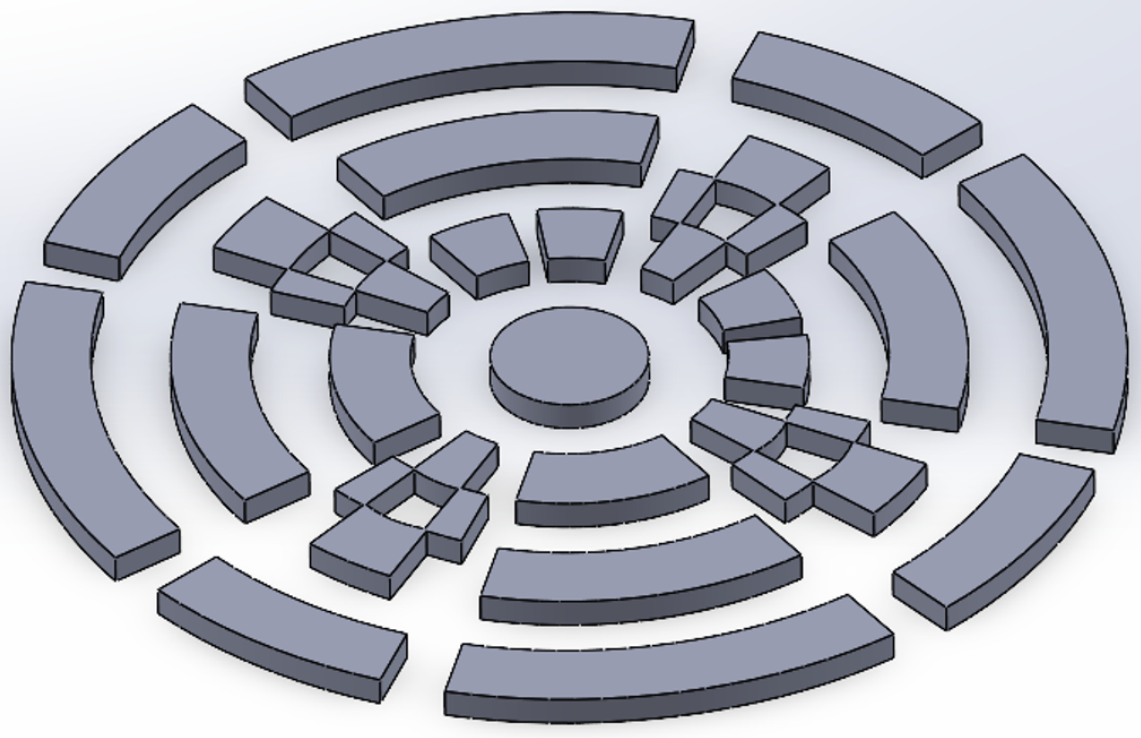}
    \caption{The view of a fractal concentric circular patch of our proposed metasurface antenna in Fig~\ref{fig:transmitarray_antenna}. 
    The FCC patch element diameter is reduced by a geometric scaling factor of 0.8 from each inner ring to the adjacent outer ring, ~\cite{sharifi2020optimization}.}
    \label{fig:FCC patches}
\end{figure}

The Random Hill-Climbing (RHC) framework has been extensively employed to improve the operating bandwidth of antennas or reflectors by optimizing row-dependent resonant elements with different geometrical dimensions,\cite{fallahi2010thin,torabi2017evolutionary,sharifi2020optimization}.
RHC model is a local optimization algorithm that iteratively maximizes an objective function. Starting from an initial random or predefined solution, it generates neighboring solutions by modifying the binary representation of the current solution and retains any candidate that improves the objective function.
This work adopts the objective function proposed in \cite{sharifi2022optimization} to generate optimized FCC patches Fig.~\ref{fig:FCC patches}. The resulting FCC patches are subsequently mapped onto the hemispherical transmitarray lens, thereby significantly enhancing the antenna bandwidth.
\begin{figure}[!t]
\centering
\includegraphics[
    width=1.02\columnwidth,
    height=6cm,
]{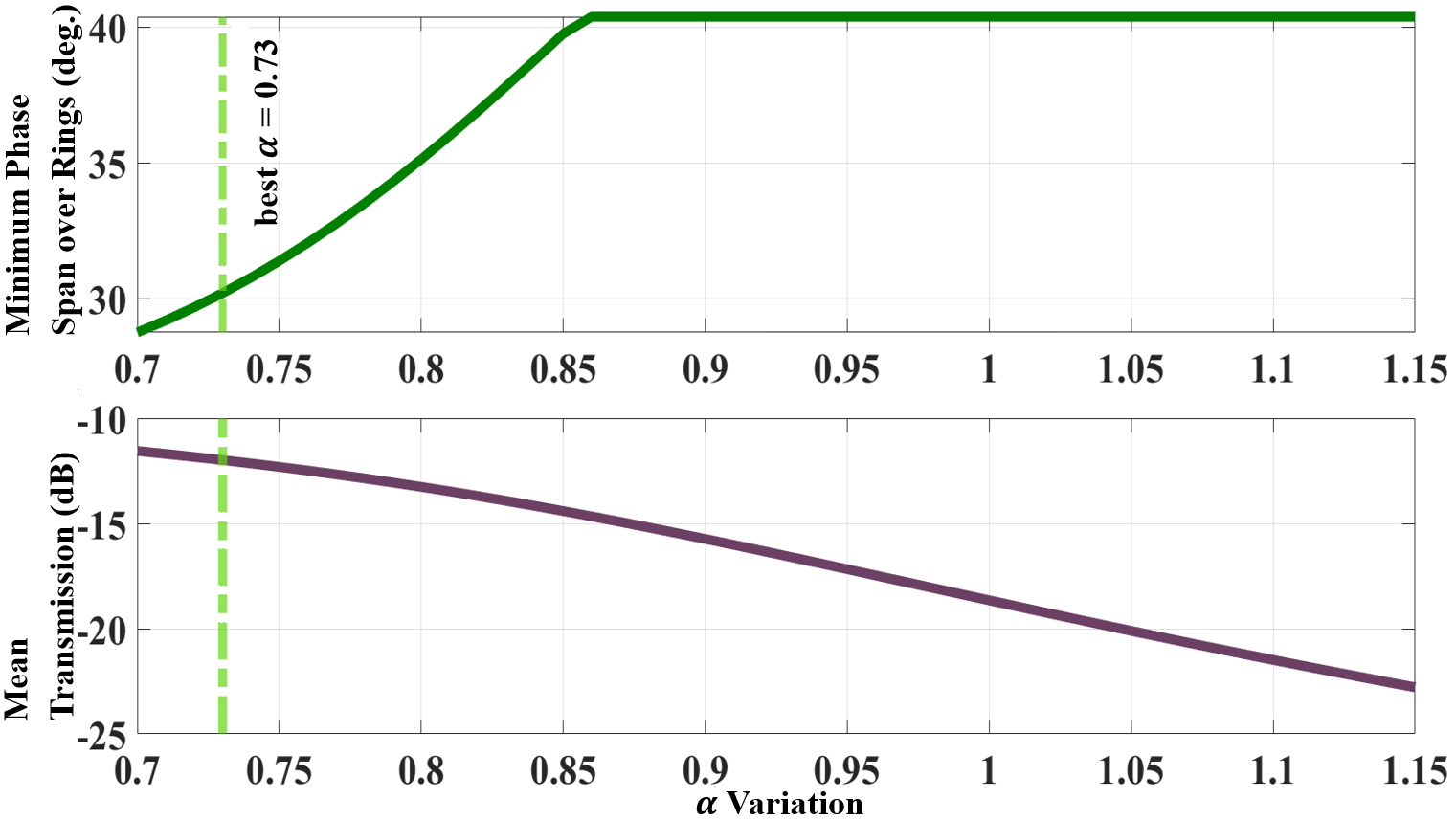}
\caption{A demonstration of the geometric scaling factor of the FCC patches was optimized, and a value of 0.8 was selected to maximize the transmission coefficient ($S_{21}$) of the unit cell.}
    \label{fig:scaling_factor}
\end{figure}
In this work, the objective is to determine the values of $\alpha$ that maximize the useful $S_{21}$ transmission-phase tuning range over the allowable chemical-potential interval 
$\mu_c = 0.1\text{--}1\,\mathrm{eV}$
while maintaining sufficiently high transmission magnitude (i.e., low insertion loss). Hence, by applying the aforementioned condition, the gold patch scaling factor per ring is determined to be $\alpha=0.8$ reasonable with a choice of $a=8$. Indeed, in our hemispherical transmitarray antenna, the dimensions of the FCC elements scale by a constant factor of 0.8 from one ring to the next, as detailed in Table I of ~\cite{komeylian2025hightechrxiv}.

\begin{figure}[!t]
\centering
\includegraphics[
    width=1.02\columnwidth,
    height=3.5cm,
]{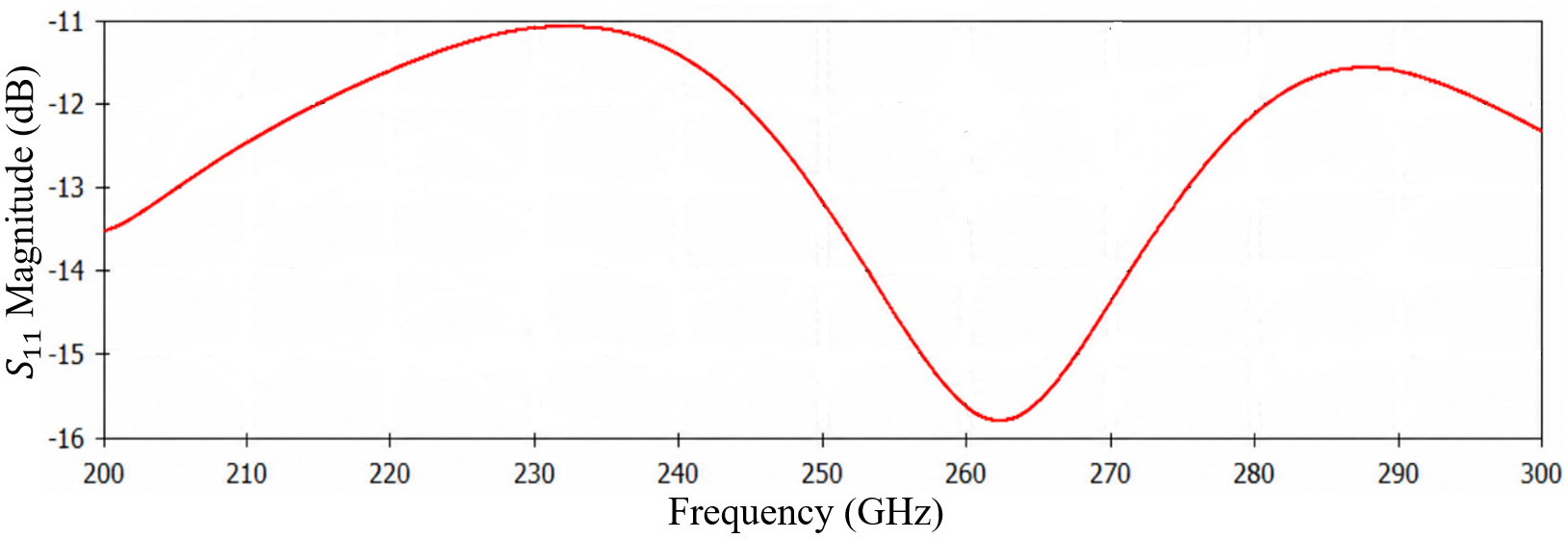}
\caption{$S_{11}$ variations in terms of frequency in GHz for our metasurface transmitarray antenna consistent with Fig.~    \ref{fig:transmitarray_antenna}, reflecting an exceptionally wide bandwidth for an operating frequency of 250 GHz and at $\theta=0^\circ$ and $\phi=0^\circ$.}
\label{fig:S11}
\end{figure}

It is evident in Fig.~\ref{fig:scaling_factor} that a geometric scaling factor of 0.8 was determined through parametric optimization to maximize the transmission coefficient of $S_{21}$. 
The unit-cell dimensions are gradually varied from one ring to the next along the radial direction of the hemispherical aperture with the scaling factor of $\alpha$. This causes each concentric ring to resonate at a distinct frequency that is closely spaced to those of its neighboring rings, resulting in overlapping resonances and a significant enhancement of the overall operating bandwidth.
The superposition of these adjacent resonances broadens the overall transmission and impedance bandwidth. 
This is noticeable in Fig.~\ref{fig:S11} that our hemispherical transmit-array antenna achieves an exceptionally wide operating bandwidth while fully preserving its beamforming capability.

\textbf{hBN$_1$}: The first hBN layer serves as ultra-thin dielectric isolation layers and gate dielectrics, preventing direct electrical contact between graphene and metallic elements while enabling efficient electrostatic tuning of graphene conductivity. The $hBN_1$ layer electrically isolates the graphene layer from the metallic patches and acts as a dielectric spacer, while the graphene layer provides voltage-controlled tunability through variation of its surface conductivity.
$hBN_1$ electrically isolates the graphene layer from the metallic FCC patches, preventing unintended short circuits while providing efficient capacitive coupling for electrostatic tuning. 

\textbf{Graphene sectors}: First, the continuous graphene layer is discretized into electrically isolated trapezoidal sectors, each corresponding to a single FCC element, thereby enabling independent voltage biasing and programmable electromagnetic control. The FCC gold patches are centrally aligned with their corresponding graphene unit cells to maximize the electromagnetic interaction between the resonant metallic structure and the tunable graphene layer. This arrangement enhances the achievable transmission-phase control while preserving independent electrical tuning through the intermediate hBN dielectric layer.
The graphene sectors introduce a dynamically reconfigurable surface impedance by varying the graphene chemical potential through applied bias voltage, thereby allowing independent tuning of the transmission phase and amplitude responses of the metasurface patches. The applied voltage bias modifies the graphene carrier concentration, which in turn dynamically adjusts its surface conductivity and provides tunable control of the electromagnetic response and transmission phase of each unit cell.

\textbf{hBN$_2$}: An alternative fabrication strategy is proposed to preserve the integrity of the $SiO_2$ substrate by avoiding direct surface etching. In this approach, the second uniform hBN layer of $hBN_2$ is first deposited on the $SiO_2$ substrate, and the required microchannels are subsequently patterned and formed exclusively within this hBN layer. 
During fabrication, microchannels are designed to enable independent voltage biasing and dynamic control of each graphene sector while maintaining the mechanical integrity and low-loss electromagnetic properties of the supporting substrate. Therefore, the microchannel fabrication process is confined to the hBN layer rather than being performed directly on the $SiO_2$ substrate, thereby preserving the inherent low-loss characteristics and high-quality electromagnetic performance of the proposed metasurface structure.

\textbf{SiO$_2$ lens}: 
The $SiO_2$ layer provides the electromagnetic interface between the horn excitation and the reconfigurable metasurface layers. It mechanically supports our conformal metasurface antenna, introduces controlled dielectric loading, and preserves the hemispherical geometry required for the radial transformation of the transmitted electromagnetic wavefront.

\textbf{Circular feedhorn}: The excitation horn is positioned at, or very near, the geometric center of the hemisphere, ensuring that the incident electromagnetic field illuminates the hemispherical aperture predominantly along the local radial direction. This configuration minimizes phase variation across the aperture and enables efficient coupling between the feed source and the conformal metasurface elements.
The horn antenna serves as the high-frequency RF excitation source, while the graphene bias network provides low-frequency electrostatic tuning. The biasing circuitry is therefore considered a control mechanism rather than an RF feeding structure and should not be included in the electromagnetic excitation path.

\subsubsection{Gold Patch Dimensions and Geometric Scaling Analysis}
The electromagnetic characteristics of our transmitarray antenna are substantially affected by the geometric distribution of gold patches across the hemispherical aperture. Our proposed active metasurface is conformal to a hemispherical lens with the radius of $R$. 
Specifically, $a$, $n$, $\alpha$ mainly govern the spatial phase sampling of the unit cells over the hemispherical aperture. 
They control the number of unit cells per ring, the local azimuthal period of $D_{(n,k),\phi}$, the effective area of each unit cell, $\Delta S_n$, and the azimuthal coordinate of each unit cell of $\phi_{(n,k)}$. 

Subsequently, these spatial dimensions determine the local graphene surface impedance of $Z_{g,(n,k),\phi}$, and the equivalent admittance of the gold-patch FSS of $Y_{Au,(n,k),\phi}$.
On the other hand, the scaling parameter of $\alpha$ regulates the dimensions of the resonant metallic structures. Specifically, the gold-patch length, $L_{Au,(n,k),\xi}$, and the corresponding inter-patch gap width, $d_{Au,(n,k),\xi}$, specifically, the gold-patch length of $L_{Au,(n,k),\xi}$, and their corresponding inter-patch gap width of $d_{Au,(n,k),\xi}$, is explicitly determined by $\alpha$. Hence, these geometrical parameters control the equivalent gold-patch FSS admittance of $Y_{\mathrm{Au},(n,k),\xi}$, which is accompanied by regulating the magnitude and phase of the local transmission coefficient of $S_{21,(n,k)}$.
Consequently, whereas $a$ and the distances between unit cells define the spatial sampling grid and unit cell distribution across the hemispherical aperture, $\alpha$ controls the local electromagnetic response by tuning the characteristics of the gold-patch FSS. 
The functional dependencies between these design parameters and the resulting electromagnetic quantities are summarized as follows,
\begin{equation}
\begin{split}
(a,\rho)
\longrightarrow
\Big\{&
N_n,\,
D_{\phi,(n,k)},\,
\Delta S_{(n,k)},\,
\phi_{(n,k)},\\
&Z_{g,(n,k),\xi},\,
Y_{\mathrm{Au},(n,k),\xi}
\Big\}
\end{split}
\end{equation}
\begin{equation}
\begin{split}
\alpha
\longrightarrow
\Big\{&
L_{\mathrm{Au},(n,k),\xi},\,
d_{\mathrm{Au},(n,k),\xi},\,
Y_{\mathrm{Au},(n,k),\xi},\,
S_{21,(n,k)}
\Big\}
\\
&\xi \in \{\theta,\phi\}
\end{split}
\end{equation}
where $\xi \in \{\theta,\phi\}$ represents the polarization index, corresponding to the $\theta$- or $\phi$-polarized electromagnetic field component.
Consequently, the local surface impedance, the local transmission coefficient, and the overall spherical array factor are formulated explicitly as functions of the design parameter vector, 
\begin{equation}
\mathbf{p}
=
\left\{
a,\,
\alpha,\,
N_r,\,
R,\,
\theta_{\max},\,
V_{n,k},\,
f,\,
\ldots
\right\}
\end{equation}
where $R$, $V_{(n,k)}$, and $f$ denote the hemisphere radius, a bias voltage applied to the $(n,k)$-th graphene sector, and the operating frequency, respectively. $\theta_{max}$ represents the maximum polar angle covered by the hemispherical aperture. 
The analytical and algorithmic optimization framework is governed by these geometrical parameters, enabling the hemispherical transmitarray antenna to realize adaptive beam steering, and thereby improving transmission efficiency.
The maximum element size is selected as
$D_{n,max}=0.5\lambda_{n,max}$, while the spacing between adjacent
elements within each row is fixed at $0.15\lambda_{n,max}$, where
$\lambda_{n,max}$ denotes the wavelength corresponding to the resonant
frequency of the $n$-th ring.
\begin{figure}[htbp]
    \centering
    \includegraphics[width=1\linewidth]{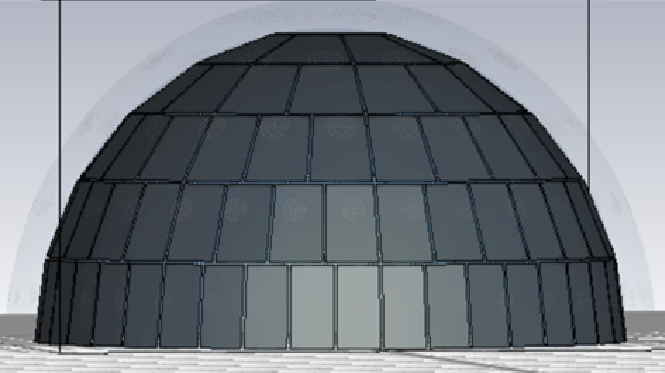}
    \caption{The hemispherical transmitarray consists of 121 graphene sectors distributed over the hemispherical surface, where each locally planar cell is approximated by a trapezoidal geometry. Each graphene sector is independently controlled by a distinct gate voltage corresponding to a graphene chemical potential ranging from 0.1 to 1.0 eV, while all cells share a common carrier relaxation time $\tau=0.1ps$ and an operating temperature of $T=300K$. All the graphene sectors are physically and electrically isolated from adjacent sectors to ensure independent electrostatic biasing and stable reconfigurable operation.}
    \label{fig:graphene-sectors}
\end{figure}
\subsubsection{Angular Discretization and Geometric Arrangement of Graphene Sectors}
Assume that a uniform graphene layer lies on a hemisphere with radius $R$. The position vector of the center of the $(n,k)$-th graphene sector,					
\begin{equation}
\mathbf{r}_{(n,k)}
=
R\left[
\sin\theta_n\cos\phi_{(n,k)}\,\hat{\mathbf{x}}
+
\sin\theta_n\sin\phi_{(n,k)}\,\hat{\mathbf{y}}
+
\cos\theta_n\,\hat{\mathbf{z}}
\right]
\end{equation}
To enable independent electromagnetic control of each FCC patch element, an individual bias voltage is assigned to each graphene sector. Hence, the continuous graphene layer is therefore discretized into electrically isolated trapezoidal sectors. Indeed, the continuous graphene layer deposited on the hemispherical surface is discretized into electrically isolated trapezoidal sectors using the angular intervals of $\Delta\theta$ and $\Delta\phi$. The dimensions of these sectors along the two orthogonal tangent directions are derived directly from the spherical coordinate system. 
From the mathematical derivation, the elevation position of $(n,k)$-th graphene sectors are given by, 
\begin{equation}
\theta_n
=
\left(n-\frac{1}{2}\right)\Delta\theta,
\qquad n\geq 1
\end{equation}
\begin{equation}
\Delta\theta
=
\frac{\theta_{\max}}{N_r}
\end{equation}
where $\theta_{max}$ denotes the maximum polar angle of the active hemispherical aperture measured from the positive $z$-axis, which is equal to $\pi/2$  for a complete hemispherical aperture. 
The azimuthal position of the $(n,k)$-th unit cell is given by, 
\begin{equation}
\phi_{(n,k)}
=
\phi_{0,n}
+
\frac{2\pi(k-1)}{N_n},
\qquad
k=1,2,\ldots,N_n
\end{equation}
where $\phi_{0,n}$ represents the initial azimuthal offset of the $(n,k)$-th ring that can be introduced to stagger adjacent rings and improve the spatial distribution of the graphene sectors. $N_r$ and $N_n$ represent the number of rings, excluding the apex unit cell, and the number of unit cells distributed on each ring, respectively.		
The local cell period along the azimuthal direction is determined by the circumferential arc length of each ring and varies with the polar position on the hemisphere by the following form, 
\begin{equation}
D_{\phi,(n,k)}
=
R\sin\!\left[ \theta_{(n,k)}\right]\Delta\phi
=
\frac{2\pi R\sin\! \left[\theta_{(n,k)}\right]}{N_{r}}
\end{equation}
For a linear ring-population model, the number of graphene sectors distributed along the $n$-th ring is expressed as, 
\begin{equation}
\label{eq:Nr}
N_r = a\left[1+\beta(n-1)\right]
\end{equation}
where $a$ represents the initial number of unit cells in the first ring and $\beta$ denotes the ring-to-ring growth coefficient. 
The element distribution for the first five rings, with $N_n={[8,16,24,32,40,48,64,80]}$, is exactly described by Eq.\ref{eq:Nr} using $a=8$ and $\beta=1$, resulting in the linear relation $N_r=8n$.

The local azimuthal period in $n$-th ring becomes, 
\begin{equation}
D_{\phi,(n,k)}(a,\beta)
=
\frac{2\pi R\sin\theta_{(n,k)}}
{a\left[1+\beta(n-1)\right]}
\end{equation}
in contrast to the constant meridional period, the azimuthal period is substantially dependent on the polar angle due to the spherical curvature.
Near the apex of the hemisphere $(\theta_n\rightarrow0)$, the circumference of each ring approaches zero, resulting in progressively smaller azimuthal cell periods and a correspondingly denser spacing of the graphene sectors.
Conversely, toward the equatorial region $(\theta_n\rightarrow\pi/2)$, the available circumference increases, requiring a larger number of sectors to maintain approximately uniform spatial sampling and conformal coverage over the hemispherical aperture.
\begin{equation}
D_{\theta,(n,k)}=R\Delta\theta
\end{equation}
where $\Delta\theta$ represents the angular spacing between consecutive rings. For a uniformly discretized hemispherical aperture, this meridional period remains constant for all rings because it is independent of the polar angle $\theta_{(n,k)}$. 
Hence, the approximate area of each sector on the hemispherical curvature is given by,
\begin{align}
\Delta S_{(n,k)}
\approx
R^2\sin\theta_n\,\Delta\theta\,
\frac{2\pi}{N_n}
=
R^2\sin\theta_n\,\Delta\theta\,\Delta\phi_n\\
\Delta\phi_n=\frac{2\pi}{N_n}
\end{align}
The graphene sectors are designed to be electrically isolated from one another, thereby enabling independent tuning of the surface impedance of each unit cell through separate gate-voltage biasing. In this sense, adjacent graphene cells should be separated by finite and physical gaps in the following expressions, 
\begin{equation}
\begin{aligned}
G_{\theta,(n,k)}
&=
D_{\theta,(n,k)}
-
g_{\theta,(n,k)}\\
G_{\phi,(n,k)}
&=
D_{\phi,(n,k)}
-
g_{\phi,(n,k)}
\end{aligned}
\end{equation}
Unlike a planar periodic array, the period in the $\phi$-direction depends on $\theta$. Near the apex of the hemisphere, $\sin\theta$ is small, and therefore the sectors become more compressed along the azimuthal direction.
\section{Voltage-Controlled Surface Impedance and Transmission Modeling of the Graphene-based Transmitarray Antenna}
The electromagnetic properties of the graphene layers are characterized by using the Kubo formulation. The graphene sheet is modeled as an infinitesimally thin conductive surface with a surface impedance defined as, \cite{dash2022active},
\begin{equation}
Z_{\mathrm{g}}(\omega,\mu_c)=\frac{1}{\sigma_{\mathrm{g}}(\omega,\mu_c)}
\end{equation}
where $\sigma_g$  represents the complex surface conductivity of the graphene. 
For the low-terahertz frequencies considered in this work, the graphene conductivity is dominated by the intraband contribution of the Kubo formulation and can therefore be approximated by the Drude model, 
\begin{equation}
\sigma_g(\omega,\mu_c)
=
\frac{\sigma_0(\mu_c)}
{1+j\omega\tau}
\end{equation}
The corresponding DC conductivity, $\sigma_0$, is given by, \cite{dash2022active},

\begin{equation}
\sigma_0
=
\frac{e^2 k_B T \tau}{\pi\hbar^2}
\left[
\frac{\mu_c}{k_B T}
+
2\ln\left(1+e^{-\frac{\mu_c}{k_B T}}\right)
\right]
\end{equation}
\noindent where $e$, $k_B$, $T$, $\hbar$, $\tau$ and $\mu_c$ represent the elementary charge, the Boltzmann constant, the absolute temperature, the reduced Planck constant, the carrier relaxation time, and the graphene chemical potential respectively.
For the biasing conditions considered in this work, where the graphene chemical potential satisfies $\mu_c \gg k_B T$, the Kubo formulation simplifies to the following Drude expression, 
\begin{equation}
\sigma_g(\omega,\mu_c)
\approx
\frac{e^2\mu_c}
{\pi\hbar^2(\tau^{-1}+j\omega)}
\end{equation}
where $1/\tau$ denotes the carrier scattering rate.
Considering the periodic distribution of graphene sectors over the hemispherical transmitarray aperture, the equivalent surface impedance is formulated by incorporating the intrinsic graphene conductivity as well as the reactive loading arising from the unit-cell geometry and inter-element coupling as follows, 
\begin{equation}
Z_g
=
\frac{D}{(D-d)\sigma_g}
-
\frac{j}{\omega C_{\mathrm{eff}}}
\label{eq:graphene_surface_impedance}
\end{equation}
where D and d refer to the unit-cell period and the graphene-sector dimension, respectively. 
The first term of Eq.\ref{eq:graphene_surface_impedance} represents the scaled material impedance of the graphene patches, while the second term accounts for the reactive contribution of the fringing electric fields across adjacent sectors. Hence, for the periodic arrangement of graphene sectors over the hemispherical transmitarray aperture, the equivalent surface impedance described in Eq.\ref{eq:graphene_surface_impedance} incorporates both the intrinsic graphene conductivity and the geometry-induced reactive effects of the periodic unit-cell structure.
The effective gap capacitance in Eq.~\ref{eq:Ceff} is approximated using the classical closed-form FSS expression, ~\cite{munk2000frequency},
\begin{equation}
C_{\mathrm{eff}}
=
\frac{\varepsilon_0(\varepsilon_r+1)D}{\pi}
\ln\!\left[
\csc\!\left(
\frac{\pi d}{2D}
\right)
\right]
\label{eq:Ceff}
\end{equation}
where $\epsilon_0$ and $\epsilon_r$ denote the vacuum permittivity and the relative permittivity of the underlying substrate, respectively.
The surface conductivity is dynamically adjusted via electrostatic biasing. Therefore, for the $(n,k)$-th unit cell, the electrically induced chemical potential $\mu_{c,(n,k)}$ is determined by the applied gate voltage $V_{(n,k)}$ according to,
\begin{equation}
\mu_{c,(n,k)}
=
\hbar v_F
\sqrt{
\frac{
\pi C_g \left|V_{(n,k)}-V_D\right|
}{e}
}
\end{equation}
where $v_F$, $C_g$, and $V_D$ represent the Fermi velocity, the gate capacitance per unit area, and the Dirac voltage, respectively. This local voltage modulation establishes a spatially tuning conductivity profile,
\begin{equation}
\sigma_{g,(n,k)}
=
\sigma_g\!\left(
\omega,\mu_{c,(n,k)}
\right)
\end{equation}
Consequently, adjusting the localized gate voltage directly controls the graphene sheet conductivity, thereby modulating the equivalent surface impedance of each individual transmitarray unit cell. This voltage-controlled impedance provides independent, reconfigurable tuning of the transmitted wave's amplitude and phase, enabling dynamic beam steering and adaptive wavefront manipulation in our proposed hemispherical graphene-based transmitarray antenna.
Accordingly, this section develops the equivalent surface impedance model for the multilayer transmitarray antenna.
We use the generalized sheet transition condition (GSTC) formulation and the Huygens surface concept, \cite{kuester2003averaged,holloway2016homogenization}, to determine the local voltage-controlled anisotropic impedance of the graphene sectors and gold patches.
This modeling approach decomposes the overall electromagnetic response of the transmitarray antenna into two complementary parts:
(1) the local unit-cell response, in which the graphene sectors exhibit tunable RLC characteristics similar to planar metasurface elements, and (2) the global hemispherical electromagnetic response is incorporated through the spherical arrangement of the unit cells, the ABCD-matrix formulation of the multilayer transmission system, and the corresponding $S_{21}$ transmission characteristics across the curved aperture.
\subsection{Voltage-Controlled Surface impedance of Graphene Sectors}
The graphene layer, consistent with Fig.~\ref{fig:graphene-sectors}, is modeled through the Kubo conductivity, $\sigma(\mu_c)$, which captures the dependence of the sheet conductivity on the chemical potential. Indeed, the tunability of graphene originates from the variation of its complex surface conductivity with the chemical potential, $\mu_c$. To incorporate this effect into the electromagnetic model, the graphene layer is represented as an equivalent conductive sheet characterized by the surface admittance $Y_s$. The relationship between the transmission coefficient and the sheet admittance of a zero-thickness conductive sheet is given by, 
\begin{equation}   
T=\frac{2}{2+Y_s Z_0}
\end{equation}   
where $Z_0$ is the free-space impedance.
Accordingly, the equivalent sheet admittance, which fully captures the bias-dependent electromagnetic response of graphene, is obtained as,
\begin{equation}   
Y_s=\frac{2(1-T)}{Z_0 T}
\end{equation}   
and the corresponding sheet impedance is given by, 
\begin{equation}   
Z_s=\frac{1}{Y_s}
\end{equation}
To achieve low-reflection transmission with full phase control, the unit cell
is modeled using the generalized sheet transition conditions (GSTCs) or the
equivalent Huygens' surface representation, \cite{kuester2003averaged,holloway2016homogenization}. The GSTC model
represents the finite-thickness multilayer unit cell by an equivalent
zero-thickness Huygens surface that preserves the same transmission and
reflection characteristics. Unlike a purely electric sheet approximation, the
Huygens model accounts for both electric and magnetic surface responses
generated by the combined interaction of the graphene sectors, gold FCC
patches, hBN dielectric layers, and the supporting silicon substrate.
For a transmissive metasurface supporting both electric and magnetic surface
responses, the corresponding boundary conditions are expressed as,
\begin{equation}   
\hat{\mathbf{n}} \times \left(\mathbf{H}_2-\mathbf{H}_1\right)
=
Y_{\mathrm{se}}\,\mathbf{E}_{\mathrm{av}}
\end{equation}
\begin{equation}   
\left(\mathbf{E}_2-\mathbf{E}_1\right)\times \hat{\mathbf{n}}
=
Z_{\mathrm{sm}}\,\mathbf{H}_{\mathrm{av}}
\end{equation}
When only the graphene sheet is considered, the metasurface response is 
represented solely by an electric surface admittance.
Since an infinitesimally thin graphene layer does not exhibit an intrinsic magnetic surface response, the equivalent magnetic sheet impedance is assumed to be zero, i.e., $Z_{sm}=0$. 
Accordingly, the GSTC formulation reduces to the electric sheet boundary 
condition, where $Y_{se}$ represents the equivalent electric sheet admittance 
of the graphene layer, consistent with graphene sectors in Fig.~\ref{fig:graphene-sectors}.
It is worth mentioning that the magnetic response does not originate from the
graphene layer alone; rather, it emerges from a combination of the electromagnetic interactions associated with the stacked multilayer unit cell, including the graphene sectors, metallic FCC patches, dielectric layers, and supporting substrate. Therefore, the equivalent electric surface admittance and magnetic surface impedance should be represented as second-order tensors to account for the anisotropic response of the stacked multilayer elements in the local spherical coordinate system. Accordingly, the electric surface admittance tensor is expressed as, 
\begin{equation}
\mathbf{Y}_{se}=
\begin{bmatrix}
Y_{\theta\theta} & 0\\
0 & Y_{\phi\phi}
\end{bmatrix}
\end{equation}
while the equivalent magnetic surface impedance tensor is given by,
\begin{equation}
\mathbf{Z}_{sm}=
\begin{bmatrix}
Z_{\theta\theta}^{m} & 0\\
0 & Z_{\phi\phi}^{m}
\end{bmatrix}
\end{equation}
For reflectionless transmission with a desired (or required) phase response of $\psi$, the transmission coefficient is defined as, 
\begin{equation}   
T=e^{j\psi}
\end{equation}
Hence, the corresponding equivalent electric and magnetic surface parameters for a reflectionless Huygens transmissive sheet with a desired phase response $\psi$ can be derived as, 
\begin{equation}   
Y_{\mathrm{se}}
=
-\frac{2j}{Z_0}
\tan\!\left(\frac{\psi}{2}\right)
\end{equation}
\begin{equation}   
Z_{\mathrm{sm}}
=
-2jZ_0
\tan\!\left(\frac{\psi}{2}\right)
\end{equation}
These last equations indicate that a purely electric sheet cannot generally provide complete transmission-phase coverage while preserving high transmission efficiency. 
In contrast, a combination of graphene, hBN, gold patch, and dielectric lens can behave as an effective Huygens-like transmitarray unit cell, enabling efficient and independent control of the transmitted wave amplitude and phase.

However, for the multilayer transmitarray unit cell, the ABCD-matrix formulation provides a more rigorous representation by explicitly modeling the cascaded electromagnetic interactions of the graphene sectors, hBN dielectric layers, gold FCC patches, and silicon substrate, thereby accurately predicting the overall transmission and reflection characteristics.
Because the unit-cell periods along the $\theta$- and $\phi$-directions differ, the local tangent-plane graphene surface impedance is inherently anisotropic. Accordingly, the graphene surface impedance tensor is expressed as,
\begin{equation}
\overline{\overline{Z}}_{g,(n,k)}
=
Z_{g,\theta}^{(n,k)}
\hat{\boldsymbol{\theta}}\hat{\boldsymbol{\theta}}
+
Z_{g,\phi}^{(n,k)}
\hat{\boldsymbol{\phi}}\hat{\boldsymbol{\phi}}
\end{equation}
Hence, the corresponding graphene surface impedance components along the meridional and azimuthal directions are expressed as, 
\begin{equation}   
Z_{g,(n,k),\theta}
=
\frac{D_{(n,k),\theta}}
     {G_{(n,k),\theta}\,\sigma_{g,(n,k)}}
-
\frac{j}{\omega C_{g,(n,k),\theta}}
\end{equation}   
The equivalent gap capacitances along the two orthogonal tangential directions are approximated using the closed-form FSS formulation as, \cite{costa2014overview}, \cite{dash2022active},
\begin{equation}   
C_{g,(n,k),\theta}
=
\frac{\varepsilon_0 \varepsilon_{\mathrm{eff}} D_{(n,k),\theta}}{\pi}
\ln\!\left[
\csc\!\left(
\frac{\pi g_{(n,k),\theta}}{2D_{(n,k),\theta}}
\right)
\right]
\end{equation}
Consequently, the explicit expressions demonstrate that the local graphene surface impedance is jointly determined by the voltage-controlled graphene conductivity, the anisotropic unit-cell geometry, and the direction-dependent gap capacitances.
\subsection{Gold-Patch Local Sheet Admittance}
The gold layer is modeled as a ring-dependent FSS, whose equivalent sheet admittance captures the electromagnetic response of the metallic pattern for each concentric ring. When the unit-cell dimensions are electrically small relative to the operating wavelength, the electromagnetic response is predominantly capacitive due to the strong fringing electric fields across the inter-patch gaps. Under this first-order approximation, the sheet admittance along the two orthogonal tangential directions is expressed as,
\begin{equation}
\label{eq:Au_admittance_theta}
Y_{\mathrm{Au},(n,k),\theta}
=
G_{\mathrm{Au},(n,k),\theta}
+
j\omega C_{\mathrm{Au},(n,k),\theta}
\end{equation}
\begin{equation}
\label{eq:Au_admittance_phi}
Y_{\mathrm{Au},(n,k),\phi}
=
G_{\mathrm{Au},(n,k),\phi}
+
j\omega C_{\mathrm{Au},(n,k),\phi}
\end{equation}
Here, $G_{Au}$  represents the conductive loss of the metallic FSS, while the direction-dependent capacitances describe the anisotropic electromagnetic coupling between adjacent gold FCC patches over the hemispherical transmitarray surface.
For the periodic gold-patch array, Fig.~\ref{fig:transmitarray_antenna}, the equivalent gap capacitances along the two orthogonal tangential directions are approximated using the well-established closed-form FSS capacitance formulation reported in \cite{costa2014overview} and \cite{dash2022active},
\begin{equation}    
C_{\mathrm{Au},(n,k),\theta}
=
\frac{\varepsilon_0 \varepsilon_{\mathrm{eff,Au}} D_{(n,k),\theta}}{\pi}
\ln\!\left[
\csc\!\left(
\frac{\pi d_{\mathrm{Au},(n,k),\theta}}
{2D_{(n,k),\theta}}
\right)
\right]
\end{equation}
A commonly used first-order approximation for the effective permittivity is given by,
\begin{equation}
\epsilon_{\mathrm{eff}}
\approx
\frac{\epsilon_{\mathrm{upper}}+\epsilon_{\mathrm{lower}}}{2}
\end{equation}
Accordingly, when the graphene sheet is located at the interface between an hBN layer and a silicon substrate, the effective permittivity can be approximated as,
\begin{equation}
\epsilon_{\mathrm{eff}}
\approx
\frac{\epsilon_{\mathrm{hBN}}+\epsilon_{\mathrm{Si}}}{2}
\end{equation}
Conversely, when the graphene sheet is sandwiched between two hBN dielectric layers, the effective permittivity reduces to, 
$\epsilon_{eff}\approx\epsilon_{hBN}$.
However, when the FCC patches are not electrically small or when the operating frequency approaches the intrinsic resonance of the FSS element, the capacitive approximation alone is no longer sufficient. In this regime, the finite current path on the metallic patches introduces an inductive response, and the interaction between the inductive and capacitive effects produces the resonant behavior of the unit cell. Consequently, the equivalent sheet admittance is more accurately represented by an RLC circuit model,
\begin{equation}
\begin{split}
Y_{\mathrm{Au},(n,k),\xi}
&=
G_{\mathrm{Au},(n,k),\xi}
+
j\omega C_{\mathrm{Au},(n,k),\xi}
+
\frac{1}
{j\omega L_{\mathrm{Au},(n,k),\xi}}
\end{split}
\end{equation}
In this model, $G_{Au}$ accounts for the finite conductivity of the gold patches, $C_{Au}$ represents the fringing-field capacitance between adjacent patches, and $L_{Au}$ models the inductive current flow along the finite metallic FCC elements. This RLC model accurately captures the resonant electromagnetic behavior of the gold FSS, thereby providing a more reliable description of the transmission amplitude and phase over a wide operating bandwidth.
\subsection{Equivalent Anisotropic Surface Impedance of the Multilayer Unit Cell}
Consider the complete transmitarray unit cell consisting of a graphene sector integrated with its corresponding multilayer stack, including the $SiO_2$, hBN dielectric layers and gold FCC patch. 
The complete unit cell is assumed to be as an equivalent electric sheet embedded in a symmetric medium. 
In this case, the impedance tensor of each multilayer stack unit cell is formulated in the spherical coordinate system and is characterized by its dependence on the applied graphene bias voltage in the following equation,  
\begin{equation}
\mathbf{Z}_{s,(n,k)}(V_g)
=
\begin{bmatrix}
Z_{s,(n,k),\theta}(V_g) & 0 \\
0 & Z_{s,(n,k),\phi}(V_g)
\end{bmatrix}
\end{equation}
in which each polarization component is derived by the exact Drude–RLC equivalent circuit in Eq.\ref{eq:surface_impedance}, 
\begin{equation}
\begin{split}
Z_{s,(n,k),\xi}(V_g)
=
R_{(n,k),\xi}(V_g)
+
j\omega L_{(n,k),\xi}(V_g)
\\
+
\frac{1}{j\omega C_{\mathrm{eff},(n,k),\xi}(V_g)}
\label{eq:surface_impedance}
\end{split}
\end{equation}
The resistance term representing the spatial dispersion component of the surface impedance in Eq.~\ref{eq:surface_impedance} is given by, 
\begin{equation}
R_{(n,k),\xi}
=
\frac{D_{\xi'}}{G_{\xi}}\,
\operatorname{Re}\!\left\{
\frac{1}{\sigma_{s,(n,k)}}
\right\}
\end{equation}
or equivalently,	
\begin{equation}
R_{(n,k),\xi}(V_g)
=
\frac{D_{\xi'}\left(1+\omega^2\tau^2\right)}
{G_{\xi}\,\sigma_{0,(n,k)}(V_g)}
\end{equation}
where 
\begin{equation}
\sigma_{0,(n,k)}(V_g)
=
\frac{e^2 \left|\mu_{c,(n,k)}(V_g)\right| \tau}
{\pi \hbar^2}
\label{eq:graphene_conductivity}
\end{equation}

Eq.~\ref{eq:graphene_conductivity} refers to the DC Drude conductivity. 
The corresponding inductance, L, is defined as, 
\begin{equation}
L_{(n,k),\xi}(V_g)
=
\tau R_{(n,k),\xi}(V_g)
\end{equation}    
where $\xi'$ denotes the orthogonal periodicity (e.g., for $R_{\theta},\qquad D_{\xi'} = D_{\phi}$).

The capacitance term of the surface impedance in Eq.\ref{eq:surface_impedance} is decomposed into two components of a constant geometric capacitance, determined by the physical dimensions of gaps between adjacent graphene sectors on the curved surface, and a variable quantum capacitance, which varies with the applied bias voltages.

\noindent Geometric capacitance: the fringing-field capacitance between adjacent patches on the curved surface of our hemispherical antenna is expressed by the following equation, 
\begin{equation}
C_{\mathrm{geo},(n,k),\xi}
=
\varepsilon_0 \varepsilon_{\mathrm{sub}}
\frac{D_{\theta}D_{\phi}}{G_{\xi}}
\end{equation}
\noindent Quantum capacitance: The quantum capacitance per unit area of the graphene sector is expressed as follows, 
\begin{equation}
C_{q,(n,k)}
=
\frac{2e^2}{\pi \hbar^2 v_f^2}
\left|\mu_{c,(n,k)}\right|
\end{equation}        
The absolute quantum capacitance is obtained by multiplying the per-unit-area capacitance by effective areas of graphene sectors, 
\begin{equation}
A = D_{\theta}D_{\phi}
\end{equation} 
yields, 
\begin{equation}
C_{q,\mathrm{abs},(n,k)}
=
\frac{2e^2 D_{\theta}D_{\phi}}
{\pi \hbar^2 v_f^2}
\left|\mu_{c,(n,k)}\right|
\end{equation}
Consequently, because the quantum capacitance acts in series with the geometric capacitance, the total effective gap capacitance is given by,
\begin{equation}
\frac{1}{C_{\mathrm{eff},(n,k),\xi}(V_g)}
=
\frac{1}{C_{\mathrm{geo},(n,k),\xi}}
+
\frac{1}{C_{q,\mathrm{abs},(n,k)}}
\end{equation}
which can be simplified as,
\begin{equation}
C_{\mathrm{eff},(n,k),\xi}(V_g)
=
\frac{
C_{\mathrm{geo},(n,k),\xi}
\,C_{q,\mathrm{abs},(n,k)}
}{
C_{\mathrm{geo},(n,k),\xi}
+
C_{q,\mathrm{abs},(n,k)}
}
\end{equation}                                
Finally, the final lumped surface impedance as a function of $V_g$ is obtained by substituting the voltage-dependent resistance, inductance, and effective capacitance into the equivalent RLC circuit and yields the complete voltage-controlled surface impedance,
\begin{equation}
\begin{split}
Z_{s,(n,k),\xi}(V_g)
&=
\frac{D_{\xi'}}{G_{\xi}\,\sigma_{0,(n,k)}}
\left(1+\omega^2\tau^2\right)
+
j\omega
\frac{\tau D_{\xi'}}{G_{\xi}\,\sigma_{0,(n,k)}}
\\
&\quad
-
\frac{j}{\omega C_{\mathrm{eff},(n,k),\xi}(V_g)}
\end{split}
\label{eq:surface_impedance_voltage}
\end{equation}
Substituting the explicit expression for the effective capacitance in Eq.~\ref{eq:surface_impedance_voltage},
\begin{equation}
C_{\mathrm{eff},(n,k),\xi}(V_g)
=
\left(
\frac{1}{C_{\mathrm{geo},(n,k),\xi}}
+
\frac{\pi \hbar^2 v_f^2}
{2e^2 D_{\theta}D_{\phi}\left|\mu_{c,(n,k)}\right|}
\right)^{-1}
\end{equation}
Substituting these constituent terms into the equivalent circuit model yields the lumped surface impedance explicitly as a function of the gate voltage,
\begin{equation}
\begin{split}
Z_{s,(n,k),\xi}(V_g)
&=
\frac{D_{\xi'}\left(1+\omega^2\tau^2\right)}
{G_{\xi}\,\sigma_{0,(n,k)}}
+
j\omega\,
\frac{\tau D_{\xi'}}
{G_{\xi}\,\sigma_{0,(n,k)}}
- \\
&\quad
\frac{j}{\omega}
\left(
\frac{1}{C_{\mathrm{geo},(n,k),\xi}}
+
\frac{\pi\hbar^2 v_f^2}
{2e^2 D_{\theta}D_{\phi}\left|\mu_{c,(n,k)}\right|}
\right)
\end{split}
\end{equation}
then the complete voltage-controlled surface impedance becomes,
\begin{equation}
\begin{split}
Z_{s,(n,k),\xi}(V_g)
&=
\underbrace{\frac{D_{\xi'}\left(1+\omega^2\tau^2\right)}
{G_{\xi}\,\sigma_{0,(n,k)}}
}_{R_{\xi}(V_g)}
+
j\omega\,
\underbrace{
\frac{\tau D_{\xi'}}
{G_{\xi}\,\sigma_{0,(n,k)}}
}_{L_{\xi}(V_g)}
\\
&\quad
-
\frac{j}{\omega}
\underbrace{
\left(
\frac{1}{C_{\mathrm{geo},(n,k),\xi}}
+
\frac{\pi\hbar^2 v_f^2}
{2e^2D_{\theta}D_{\phi}
\left|\mu_{c,(n,k)}\right|}
\right)
}_{\displaystyle \frac{1}{C_{\mathrm{eff},\xi}(V_g)}}
\end{split}
\label{eq:voltage_controlled_surface_impedance}
\end{equation}
or, equivalently,
\begin{equation}
\begin{split}
Z_{s,(n,k),\xi}(V_g)
&=
R_{(n,k),\xi}(V_g)
+
j\omega L_{(n,k),\xi}(V_g)
\\
&\quad
-
\frac{j}{\omega C_{\mathrm{eff},(n,k),\xi}(V_g)}
\end{split}
\end{equation}	
This formulation rigorously accounts for graphene geometry, substrate permittivity, carrier scattering time, kinetic inductance, and voltage-dependent quantum capacitance. Consequently, the surface impedance of each graphene sector is directly modulated by the applied gate voltage of $V_g$ via its influence on the graphene chemical potential. This dependency establishes the physical framework for a tunable electromagnetic response, enabling dynamic phase control and adaptive beam steering in our hemispherical transmitarray antenna.
\subsection{Equivalent Lumped RLC Circuit model of Proposed Hemispherical Transmitarray Unit Cell}
The complete voltage-controlled impedance model of the multilayer transmitarray unit cell, which incorporates the electromagnetic loading effects of the graphene sectors and surrounding dielectric layers, Fig.~\ref{fig:transmitarray_antenna}, is formulated as, 
\begin{equation}
Z_{\mathrm{cell}}(V_g)
=
Z_g(V_g)
+
Z_{\mathrm{hBN}_2}
+
Z_{\mathrm{SiO}_2}
+
\frac{1}{j\omega C_g}
\label{eq:cell_impedance}
\end{equation}
where $Z_g$ $(V_g)$, which is identical to $R_{s,(n,k),\xi}(V_g)$ in Eq.\ref{eq:voltage_controlled_surface_impedance} denotes the voltage-dependent graphene sheet impedance, while $Z_{(hBN_2)}$ and $Z_{(SiO_2)}$ represent the impedance contributions associated with the finite-thickness dielectric layers. The effective capacitance of $C_{eff,(n,k),\xi}(V_g)$ in Eq.\ref{eq:voltage_controlled_surface_impedance}, which is equivalent to $C_g$ in Eq.\ref{eq:cell_impedance}, represents the direction-dependent gap capacitance arising from the fringing fields between neighboring graphene sectors along the $\xi$-polarized direction. By substituting the equivalent graphene RLC model and the dielectric capacitive impedances, the total unit-cell impedance becomes,
\begin{equation}
\begin{split}
Z_{\mathrm{cell}}(V_g)
&=
R_g(V_g)
+
j\omega L_g(V_g)
\\
&\quad
+
\frac{1}{j\omega C_{\mathrm{hBN}_2}}
+
\frac{1}{j\omega C_{\mathrm{SiO}_2}}
+
\frac{1}{j\omega C_g}
\end{split}
\end{equation}
where $L_g$ $(V_g )$, which is identical to $L_{(n,k),\xi} (V_g)$ in Eq.\ref{eq:voltage_controlled_surface_impedance}, denotes the inductance associated with the graphene sheet impedance.
Accordingly, the impedance contributions of the $SiO_2$ substrate and upper hBN dielectric layer are given by, 
\begin{equation}
Z_{\mathrm{SiO}_2}
=
j\,\frac{\eta_0}{\sqrt{\varepsilon_{r,\mathrm{SiO}_2}}}
\tan\!\left(
k_0\sqrt{\varepsilon_{r,\mathrm{SiO}_2}}\,
t_{\mathrm{SiO}_2}
\right)
\end{equation}
and
\begin{equation}
Z_{\mathrm{hBN}_2}
=
j\,\frac{\eta_0}{\sqrt{\varepsilon_{r,\mathrm{hBN}}}}
\tan\!\left(
k_0\sqrt{\varepsilon_{r,\mathrm{hBN}}}\,
t_{\mathrm{hBN}_2}
\right)
\end{equation}
where $\eta_0$ and $k_0$ refer to the free-space impedance and the free-space wavenumber, respectively. $t_{(SiO_2)}$ and $t_{(hBN_2)}$ denote the thicknesses of the corresponding dielectric layers. 
Consequently, the applied gate voltage modifies the graphene chemical potential and conductivity, thereby dynamically tuning the effective impedance of each transmitarray unit cell while the dielectric contributions remain constant, \cite{yadav2025surface,dmitriev2023multifunctional,dash2022active}.

It is worth mentioning that the capacitance of the $hBN_1$ dielectric layer is not included in the equivalent surface impedance model because it remains constant and is independent of the applied bias voltage. Since this layer does not contribute to the voltage-dependent tuning mechanism, its electromagnetic effect is incorporated into the fixed dielectric loading of the multilayer structure rather than represented as an independent lumped circuit element. Consequently, only the voltage-dependent components are retained in the equivalent circuit formulation.
\begin{figure}[htbp]
    \centering    \includegraphics[width=1\columnwidth,height=5cm,keepaspectratio]{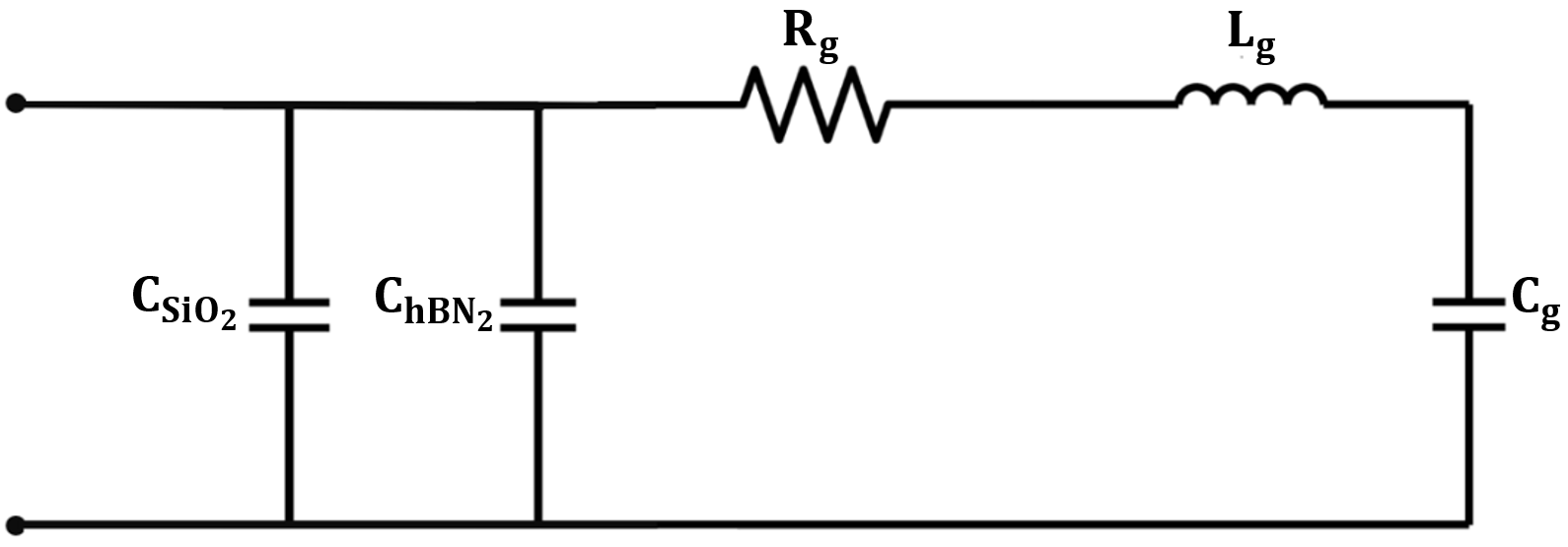}
    \caption{Equivalent circuit model for the proposed hemispherical graphene transmitarray unit cell, including the dielectric capacitances of the $SiO_2$, $C_{SiO_2}$, and upper hBN layer, $C_{hBN_2}$. The graphene sheet impedance modeled by the series resistance of $R_g$ and inductance of $L_g$, and the gap capacitance of $C_g$. The lower hBN layer is excluded from the circuit model because it remains invariant and acts as a constant parameter.}
    \label{fig:RLC Circuit}
\end{figure}
\section{Local ABCD model and Voltage-Controlled Transmission Coefficient}
The ABCD-matrix formulation provides a rigorous analytical framework for evaluating $S_{21}$ of the proposed multilayer transmitarray antenna by accounting for the cascaded electromagnetic response, interlayer coupling, dielectric propagation, and tunable graphene surface impedance.
The analytical framework accurately captures the cascaded electromagnetic response, finite-thickness effects of individual layers, and impedance transformations occurring throughout the multilayer structure. Furthermore, it accounts for the multiple internal reflections at the interfaces among the constituent material layers, including the graphene sectors, hBN dielectric layers, gold FCC patches, and silicon substrate.
For a dielectric layer under normal incidence, the corresponding ABCD matrix can be generally expressed as,
\begin{equation}
\mathbf{P}_i
=
\begin{bmatrix}
\cos\!\left(k_i d_i\right)
&
j Z_i \sin\!\left(k_i d_i\right)
\\
j Z_i^{-1}\sin\!\left(k_i d_i\right)
&
\cos\!\left(k_i d_i\right)
\end{bmatrix}
\end{equation}
where $k_i$  and $Z_i$  denote the propagation constant and characteristic impedance of the $i$-th dielectric layer, respectively, given by,
\begin{equation}
k_i = k_0\sqrt{\varepsilon_{r,i}}
\end{equation}
\begin{equation}
Z_i = \frac{\eta_0}{\sqrt{\varepsilon_{r,i}}}
\end{equation}
where $\eta_0$, and $\epsilon_{r,i}$ represent the free-space impedance, and relative permittivity of the $i$-th dielectric layer, respectively. 
An equation of \(Y_d=\frac{1}{Z_d}\) represents the characteristic admittance of the dielectric layer.
The polarization-dependent graphene sheet admittance of the $(n,k)$-th unit cell is obtained by inverting the corresponding element of the anisotropic graphene impedance tensor, expressed as,
\begin{equation}
Y_{g,(n,k),\xi}
=
\left(
Z_{g,(n,k),\xi}
\right)^{-1}
\end{equation}
where \(Z_{g,(n,k),\xi}\) represents the graphene surface impedance component associated with the $\xi$-polarized electromagnetic response of the unit cell.
Using the equivalent transmission-line representation, the graphene sheet is modeled as a shunt admittance and described by the following ABCD matrix, 
\begin{equation}
\mathbf{S}_{g,(n,k),\xi}
=
\begin{bmatrix}
1 & 0 \\
Y_{g,(n,k),\xi} & 1
\end{bmatrix}
\end{equation}
Similarly, the gold FSS layer is modeled as an equivalent shunt sheet admittance, with its corresponding ABCD matrix expressed as follows, 
\begin{equation}
\mathbf{S}_{\mathrm{Au},(n,k),\xi}
=
\begin{bmatrix}
1 & 0 \\
Y_{\mathrm{Au},(n,k),\xi} & 1
\end{bmatrix}
\end{equation}
where \(Y_{\mathrm{Au},(n,k),\xi}\) denotes the polarization-dependent sheet admittance of the gold FSS layer along the selected polarization direction of $\xi$. 

By incorporating the dielectric propagation layers, the complete transfer matrix of the $(n,k)$-th unit cell is obtained by cascading the individual layer matrices as, 
\begin{equation}
\mathbf{m}_{(n,k),\xi}
=
\mathbf{P}_{\mathrm{Si}}\,
\mathbf{P}_{\mathrm{hBN_1}}\,
\mathbf{S}_{g,(n,k),\xi}\,
\mathbf{P}_{\mathrm{hBN_2}}\,
\mathbf{S}_{\mathrm{Au},(n,k),\xi}
\end{equation}
where $\mathbf{P}_\mathrm{Si}$, $\mathbf{P}_{\mathrm{hBN_1}}$, and $\mathbf{P}_{\mathrm{hBN_2}}$ represent the dielectric propagation matrices of the silicon and hBN layers, respectively, accounting for the phase accumulation and impedance transformation through the corresponding dielectric regions.
The resulting ABCD transfer matrix of the $(n,k)$-th unit cell can be written as,
\begin{equation}
\mathbf{m}_{(n,k),\xi}
=
\begin{bmatrix}
A_{(n,k),\xi} & B_{(n,k),\xi} \\
C_{(n,k),\xi} & D_{(n,k),\xi}
\end{bmatrix}
\end{equation}
Accordingly, the local transmission coefficient of the graphene-loaded $(n,k)$-th unit cell is obtained from the ABCD parameters as, 
\begin{equation}
\begin{split}
S_{21,(n,k),\xi}
&=
\\
&\quad
\frac{2}{
A_{(n,k),\xi}
+
\frac{B_{(n,k),\xi}}{Z_{\mathrm{out}}}
+
Z_{\mathrm{in}}\,C_{(n,k),\xi}
+
\frac{Z_{\mathrm{in}}}{Z_{\mathrm{out}}}\,
D_{(n,k),\xi}
}
\end{split}
\end{equation}
where $Z_{in}$ and $Z_{out}$  denote the characteristic impedances of the media surrounding the unit cell, respectively.
\section{Analytical and Practical mapping from the Desired Beam Direction to the Bias Voltages of Graphene Sectors}
Assuming a horn antenna is positioned at the geometric center of the hemispherical lens, the incident electromagnetic wave propagates radially and illuminates the conformal metasurface aperture with an approximately uniform phase distribution. 
The objective is to synthesize a transmitted beam in the desired direction, 
\begin{equation}
\hat{\mathbf{s}}_0
=
\left(
\sin\theta_0\cos\phi_0,\,
\sin\theta_0\sin\phi_0,\,
\cos\theta_0
\right)
\label{eq:unit_vector}
\end{equation}
thereby supporting adaptive beam steering and continuous tracking of moving targets.
To achieve beam steering toward the desired direction $\hat{\mathbf{s}}_0$ under spherical-wave excitation from the feedhorn, the required transmission phase response of the multilayer stack $(n,k)$-th unit cell on the conformal hemispherical transmitarray aperture is formulated as follows,
\begin{equation}
\psi_{(n,k)}^{\mathrm{req}}
=
-k_0\,\hat{\mathbf{s}}_0 \cdot \mathbf{r}_{(n,k)}
-
\angle E_{\mathrm{inc}}^{\mathrm{horn}}\!\left(\mathbf{r}_{(n,k)}\right)
+
C
\end{equation}
where $k_0$  and $\mathbf{r}_{(n,k)}$  denote the free-space wavenumber and the position vector of the $(n,k)$-th unit cell on the hemispherical aperture, respectively. A constant of $C$ represents an arbitrary constant phase reference.
We first establish a four-step analytical framework for determining the bias voltages of the graphene sectors corresponding to desired beam directions of moving targets. The practical realization of this framework is then presented as an eight-step voltage-controlled beam-steering synthesis procedure in Table~\ref{tab:voltage-controlled-synthesis}.

\textbf{Step 1}: Analytical derivation of the required transmission phase distribution: since the feedhorn phase center coincides with the geometric center of our hemispherical lens, all unit cells on the conformal aperture are approximately equidistant from the feedhorn. Consequently, the incident phase distribution across the hemispherical aperture is nearly uniform and can be expressed as,
\begin{equation}
\angle E_{\mathrm{inc}}^{\mathrm{horn}}
\!\left(\mathbf{r}_{(n,k)}\right)
\approx -k_0 R
\end{equation}
where R refers to the hemisphere radius. Since this propagation phase term is identical for all unit cells, it can be incorporated into the constant phase reference of $C$, yielding,
\begin{equation}
\psi_{(n,k)}^{\mathrm{req}}
\approx
-k_0\,\hat{\mathbf{s}}\cdot\mathbf{r}_{(n,k)}
+
C
\label{eq:required_phase_distribution}
\end{equation}
The dot product between the desired beam direction vector $s_0$  and the position vector $\mathbf{r}_{(n,k)}$ of the $(n,k)$-th unit cell on the hemispherical aperture, which is expressed as,
\begin{equation}
\hat{\mathbf{s}}\cdot\mathbf{r}_{(n,k)}
=
R\left[
\sin\Theta_0\,\sin\theta_n\,
\cos\!\left(\Phi_0-\phi_{(n,k)}\right)
\\
+
\cos\Theta_0\,\cos\theta_n
\right]
\label{eq:spatial_phase_projection}
\end{equation}
Substituting the Eq.\ref{eq:spatial_phase_projection} into the required phase formulation in Eq.\ref{eq:required_phase_distribution} yields,
\begin{multline}
\psi_{(n,k)}^{\mathrm{req}}
=
-k_0 R
\bigg[
\sin\Theta_0\,\sin\theta_n\,
\cos\!\left(\Phi_0-\phi_{(n,k)}\right)
\\
+
\cos\Theta_0\,\cos\theta_n
\bigg]
+
C
\label{eq:required_phase_hemisphere}
\end{multline}
\textbf{Step 2}: Required local transmission coefficient: assuming ideal phase-only modulation with a transmission magnitude approximately equal to unity, the required complex transmission coefficient of the stacked multilayer $(n,k)$-th unit cell is expressed as,
\begin{equation}
T_{(n,k)}^{\mathrm{req}}
=
e^{j\Phi_{(n,k)}^{\mathrm{req}}}
\label{eq:required_transmission_phase}
\end{equation}
where $\Phi_{(n,k)}^{\mathrm{req}}$ denotes the required transmission phase obtained from the beam-steering phase synthesis. This general formulation can also account for amplitude variations by incorporating the desired transmission magnitude into the complex transmission coefficient, i.e., 
\begin{equation}
T_{(n,k)}^{\mathrm{req}}
=
\left|T_{(n,k)}^{\mathrm{req}}\right|
e^{j\Phi_{(n,k)}^{\mathrm{req}}}
\end{equation} where $\left|T_{\mathrm{req}}^{(n,k)}\right| \leq 1$.

\textbf{Step 3}: Analytical mapping from required transmission coefficient to graphene surface impedance: the required transmission coefficient of each unit cell is analytically transformed into an equivalent surface impedance, providing the essential bridge between the synthesized electromagnetic response and the voltage-controlled physical implementation of the graphene metasurface, 
target beam direction (or desired direction),
\begin{equation}
\psi_{(n,k)}^{\mathrm{req}}
\;\longrightarrow\;
T_{(n,k)}^{\mathrm{req}}
\;\longrightarrow\;
Z_{s,(n,k)}^{\mathrm{req}}
\end{equation}
For a thin conductive metasurface represented by an equivalent surface impedance sheet, the complex transmission coefficient of the multilayer stack is directly determined by the effective surface impedance according to,
\begin{equation}
T=\frac{2Z_s}{2Z_s+Z_0}
\end{equation}
where $Z_s$ represents the voltage-dependent surface impedance introduced by the graphene-loaded unit cell. 
This relation establishes the analytical link between the desired or required transmission response and the physical graphene conductivity modulation.
If the stacked multilayer $(n,k)$-th unit cell is represented by an equivalent series surface impedance, the required equivalent surface impedance is obtained from the desired complex transmission coefficient as, 
\begin{equation}
Z_{s,(n,k)}^{\mathrm{req}}
=
2Z_0
\left(
\frac{1}{T_{(n,k)}^{\mathrm{req}}}
-1
\right)
\label{eq:required_surface_impedance}
\end{equation}
This expression establishes the direct mapping between the synthesized transmission response and the required local surface impedance, thereby defining the electromagnetic design target for the realization of voltage-controlled graphene unit cells.
Substituting Eq.\ref{eq:required_transmission_phase} into Eq.\ref{eq:required_surface_impedance} yields the required local surface impedance for the $(n,k)$-th graphene-loaded unit cell, expressed as follows,
\begin{equation}
\label{eq:required_surface_impedance2}
Z_{s,(n,k)}^{\mathrm{req}}
=
2Z_0
\left(
e^{-j\Phi_{(n,k)}^{\mathrm{req}}}
-1
\right)
\end{equation}
Eq.\ref{eq:required_surface_impedance2} establishes the analytical mapping between the desired beam-steering phase profile and the required local surface impedance, which is subsequently synthesized through voltage-controlled modulation of graphene conductivity.

\textbf{Step 4}: Voltage-controlled impedance realization:
the required or desired phase profile is then transformed into the corresponding local surface impedance $Z_s^{\mathrm{req}}$, which is physically synthesized through the voltage-controlled graphene conductivity $\sigma_g(\mu_c)$. Finally, the required chemical potential is converted into the corresponding gate voltage $V_g$, enabling dynamic realization of the desired beam direction in the following path, 
\begin{equation}
\psi_{(n,k)}^{\mathrm{req}}
\;\longrightarrow\;
T_{(n,k)}^{\mathrm{req}}
\;\longrightarrow\;
Z_{s,(n,k)}^{\mathrm{req}}
\;\longrightarrow\;
\mu_{c,(n,k)}
\;\longrightarrow\;
V_{g,(n,k)}
\end{equation}
Therefore, the required gate voltage of $V_{g,(n,k)}$ corresponding to the $(n,k)$-th unit cell is determined by solving the following impedance-matching equation,
\begin{equation}
Z_{s,(n,k)}
\!\left(V_{g,(n,k)}\right)
=
2Z_0
\left[
e^{-j\Phi_{(n,k)}^{\mathrm{req}}(\theta_0,\phi_0)}
-1
\right]
\end{equation}
where $Z_{s,(n,k)}$  $(V_{g,(n,k)})$ represents the voltage-dependent surface impedance of the graphene-loaded unit cell. 
Substituting the graphene impedance expression given in Eq.\ref{eq:voltage_controlled_surface_impedance} into Eg.\ref{eq:required_surface_impedance2} and separating the resulting complex equation into its real and imaginary components yields the following coupled equations, 
\begin{equation}
\operatorname{Re}\!\left\{
Z_{s,(n,k)}(\mu_c)
\right\}
=
2Z_0
\left(1-
(\cos\Phi_{(n,k)}^{\mathrm{req}})^2
\right)
\label{eq:real_surface_impedance}
\end{equation}

\begin{equation}
\operatorname{Im}\!\left\{
Z_{s,(n,k)}(\mu_c)
\right\}
=
-2Z_0\sin\Phi_{(n,k)}^{\mathrm{req}}
\label{eq:imag_surface_impedance}
\end{equation}
Accordingly, for the graphene-loaded RLC equivalent model illustrated in Fig.~\ref{fig:RLC Circuit}, the real and imaginary impedance-matching equations are expanded into the following expressions,
\begin{equation}
\frac{D_{\xi'}}{G_{\xi}\,\sigma_0(\mu_c)}
\left(1+\omega^2\tau^2\right)
=
2Z_0
\left(1-
(\cos\Phi_{(n,k)}^{\mathrm{req}})^2
\right)
\end{equation}

\begin{equation}
\omega\tau\,
\frac{D_{\xi'}}{G_{\xi}\,\sigma_0(\mu_c)}
-
\frac{1}{\omega C_{\mathrm{eff},\xi}(\mu_c)}
=
-2Z_0\sin\Phi_{(n,k)}^{\mathrm{req}}
\end{equation}
where the graphene Drude conductivity term is given by,
\begin{equation}
\sigma_0(\mu_c)
=
\frac{e^2 \mu_c \tau}{\pi \hbar^2}
\end{equation}
The effective capacitance accounting for both geometric capacitance and graphene quantum capacitance contributions is expressed as follows,
\begin{equation}
C_{\mathrm{eff},\xi}(\mu_c)
=
\left(
\frac{1}{C_{\mathrm{geo},\xi}}
+
\frac{\pi \hbar^2 v_f^2}
{2e^2 D_{\theta}D_{\phi}\left|\mu_c\right|}
\right)^{-1}
\end{equation}

By solving the coupled nonlinear equations given in \ref{eq:real_surface_impedance} and Eq.\ref{eq:imag_surface_impedance}, the required spatial distribution of graphene chemical potential of $\mu_{c,(n,k)}$ is analytically determined. The resulting chemical-potential distribution is then transformed into the corresponding gate-voltage map of $V_{g,(n,k)}$, establishing a direct electrical control mechanism for real-time programmable beam steering of the hemispherical transmitarray antenna.

Because the resulting impedance-matching equations, Eqs.\ref{eq:real_surface_impedance} and \ref{eq:imag_surface_impedance}, are transcendental due to the nonlinear dependence of graphene conductivity, quantum capacitance, and effective surface impedance on the chemical potential, a closed-form analytical solution for $V_{g,(n,k)}$ is generally not available. 
Therefore, a practical numerical optimization framework in Table~\ref{tab:voltage-controlled-synthesis} is developed to determine the voltage bias distribution of individual graphene sectors, enabling the physical realization of the synthesized electromagnetic phase profile for programmable beam steering.
In this case, the required gate voltage for each graphene unit cell is obtained by solving the nonlinear inverse impedance-matching problem. The complex residual function is defined by,
\begin{equation}
F(V_g)=Z_s(V_g)-Z_{s}^{\mathrm{req}}=0
\end{equation}
the solution is computed iteratively using the Newton--Raphson method, \cite{Jin2015},
\begin{equation}
V_g^{(i+1)}
=
V_g^{(i)}
-
\frac{F\!\left(V_g^{(i)}\right)}
{F'\!\left(V_g^{(i)}\right)}
\end{equation}
Since $V_g$ is a real-valued control variable while $Z_s$ is complex, convergence is achieved by minimizing the complex impedance residual,
\begin{equation}
\mathcal{E}(V_g)
=
\left|Z_s(V_g)-Z_{s}^{\mathrm{req}}\right|^2
\rightarrow 0
\end{equation}
This procedure is repeated for every unit cell to determine the complete gate-voltage distribution required to realize the desired phase profile across the graphene transmitarray.
\begin{table}[t]
\centering
\scriptsize
\renewcommand{\arraystretch}{0.85}
\setlength{\tabcolsep}{2pt}

\caption{Practical steps for mapping the desired beam direction to the bias voltages of the graphene sectors.}
\label{tab:voltage-controlled-synthesis}

\begin{tabularx}{\columnwidth}{
L{0.07\columnwidth}
L{0.22\columnwidth}
Y
L{0.18\columnwidth}}

\toprule
\textbf{Step} &
\textbf{Flowchart Block} &
\textbf{Mathematical and Operation Formulation} &
\textbf{Resulting Output}\\
\midrule

Step 1 &
Beam Direction Specification &
Define the desired beam-steering direction $\left(\theta_0,\phi_0\right)$ and the corresponding unit direction vector from Eq.~\ref{eq:unit_vector} &
Target beam direction for performing steering \\

\addlinespace

Step 2 &
Required Transmission Phase Synthesis &
Calculate the required phase distribution over all unit cells on the hemispherical aperture from Eq.\ref{eq:required_phase_hemisphere} &
$\Phi_{(n,k)}^{\mathrm{req}}$ \\

\addlinespace

Step 3 &
Transmission Coefficient Synthesis &
Convert the required phase profile into the desired complex transmission coefficient of Eq.\ref{eq:required_transmission_phase} &
Desired transmission response of each unit cell, $T_{(n,k)}^{\mathrm{req}}$ \\

\addlinespace

Step 4 &
Required Surface Impedance Extraction &
Map the transmission coefficient to the required local surface impedance using Eq.~\ref{eq:required_surface_impedance2} &
Required surface-impedance distribution, $Z_{s,(n,k)}^{\mathrm{req}}$ \\

\addlinespace

Step 5 &
Graphene Impedance--Voltage Mapping &
Calculate the voltage-dependent impedance relation of each graphene-loaded unit cell, $Z_{s,(n,k)}(V_g)$, using,
\begin{itemize}[leftmargin=*,nosep]
\item full-wave electromagnetic simulations
\item an analytical graphene RLC impedance model
\item a pre-computed impedance look-up table (LUT), $V_g\mapsto Z_s$
\end{itemize}
&
Voltage-dependent impedance model \\

\addlinespace

Step 6 &
Voltage Optimization for Each Unit Cell &
For each graphene sector, determine the gate voltage by solving the impedance-matching condition,
\begin{equation}
\label{eq:voltage_solution}
Z_{s,(n,k)}(V_g)-Z_{s,(s,k)}^{\mathrm{req}}\rightarrow 0
\end{equation} 
&
Voltage solution for each unit cell \\

\addlinespace

Step 7 &
Numerical Root-Finding Algorithm &
Optimize the gate voltage using Newton–Raphson or
numerical minimization,
\begin{equation}
\begin{gathered}
V_{g,(n,k)}=\\
\underset{V_g}{\arg\min}\,
\left\lVert 
Z_{s,(n,k)}(V_g)
-
Z_{s,(n,k)}^{\mathrm{req}}
\right\rVert_2
\end{gathered}
\end{equation}
&
Minimum error impedance of $\operatorname{Error}(V_g)=Z_{s,(n,k)}(V_g)-Z_{s,(n,k)}^{\mathrm{req}}$ \\

\addlinespace

Final Output &
Programmable Graphene Bias Map &
Apply the obtained voltage distribution to electrically
isolated graphene sectors, 
\begin{equation}
\left\{
V_g^{(1,1)},V_g^{(1,2)},\ldots,
V_g^{(N_n,N_r)}
\right\}
\end{equation}
&
Real-time programmable beam steering and moving-target tracking\\

\bottomrule
\end{tabularx}

\end{table}

\section{Conformal Spherical Array Factor for the Graphene-based Transmitarray Hemispherical Antenna}
The array factor is a well-established concept in antenna theory that is rigorously formulated for passive antenna arrays.
Integrating active components, such as amplifiers, transistors, or dynamically tunable elements, extends this framework beyond its conventional scope and complicates its theoretical formulation. For instance, active elements introduce spatially varying gain and phase responses that no longer satisfy the fundamental assumptions of the pattern multiplication theorem, which traditionally defines the total radiation pattern as the product of the isolated element pattern and the array factor.
Accordingly, as a first step toward extending the classical array factor formulation developed for passive antenna arrays to our hemispherical transmitarray antenna, the dynamic tunability of the unit cells is incorporated into the analytical framework.
Although our hemispherical antenna does not contain conventional active RF devices, it can be categorized as a semi-active transmitarray antenna because its radiation characteristics are governed by externally applied bias voltages. These applied voltages dynamically modulate the surface conductivity of the graphene sections, thereby tuning the overall performance.
The most rigorous formulation is to merge the conventional free-space hemispherical array factor with the graphene transmission coefficient of $S_{21,(n,k)}$  and the horn illumination phase into a single conformal array factor. 

Hence, the conformal array factor of our hemispherical transmit-array antenna is explicitly expanded in terms of excitation weights of each graphene sectors in the following formula, 
\begin{equation}
AF(\Theta,\Phi;\mathbf{p})
=
W_0 e^{j k_0 \mathbf{s}\cdot\mathbf{r}_0}
+
\sum_{n=1}^{N_r}
\sum_{k=1}^{N_n}
W_{n,k}\,
e^{j k_0 \mathbf{s}(\Theta,\Phi)\cdot\mathbf{r}_{(n,k)}}
\label{eq:array_factor}
\end{equation}
where $N_r$ and $N_n$ represent the total number of concentric rings and the number of unit cells in the $n$-th ring, respectively. $W_0$ denotes the apex excitation weight associated with the central unit cell of the hemispherical aperture. The complex excitation coefficient of each $(n,k)$-th unit cell is defined as follows, 
\begin{equation}
W_{(n,k)}
=
\Delta S_n\,
F_{\mathrm{horn}}\!\left(\theta_n,\phi_{(n,k)}\right)
F_{\mathrm{cell},(n,k)}\!\left(\Theta,\Phi\right)
S_{21,(n,k)}
\end{equation}
where $F_{horn}$  represents the incident horn illumination distribution over the hemispherical aperture. 
The individual unit-cell radiation pattern is denoted by $F_{cell,(n,k)}$. 
In the initial optimization stage, the unit-cell radiation pattern is approximated as isotropic, i.e., $F_{\mathrm{cell},(n,k)}(\Theta,\Phi)\approx 1$. This approximation significantly reduces the computational burden while retaining the dominant contributions of the hemispherical aperture geometry, feedhorn illumination, and voltage-controlled transmission characteristics.
$S_{21,(n,k)}$ represents the voltage-controlled complex transmission coefficient of the graphene-loaded unit cell, defining the amplitude and phase response of the transmitted electromagnetic field as a function of the applied bias voltage. 

Hence, the total far-field phase contribution from each unit cell is composed of three components: (1) the voltage-controlled transmission phase provided by the graphene sectors, (2) the geometrical propagation phase determined by the unit-cell position on the hemispherical aperture, and (3) the incident phase of the spherical wave generated by the feedhorn. Indeed, the total far-field phase is given by the superposition of the following three phase contributions,
\begin{equation}
\begin{split}
\Phi_{\mathrm{total},(n,k)}^{\mathrm{far-field}}
=&\;
\underbrace{\Phi_{\mathrm{trans},(n,k)}}%
_{\text{graphene-controlled transmission phase}}
\\
&\;+
\underbrace{k_0\,\hat{\mathbf{s}}\cdot\mathbf{r}_{(n,k)}}%
_{\text{geometrical propagation phase}}
+
\underbrace{
\angle E_{\mathrm{inc}}^{\mathrm{horn}}
\!\left(\mathbf{r}_{(n,k)}\right)
}_{\text{horn illumination phase}}
\end{split}
\end{equation}
Hence, the phase contribution in the array factor in Eq.\ref{eq:array_factor} can be explicitly expressed as,
\begin{multline}
e^{j k_0 \hat{\mathbf{s}}(\Theta,\Phi)\cdot\mathbf{r}_{(n,k)}}
=
e^{-j\Phi_{(n,k)}^{\mathrm{desired}}}
\times
\\
\exp\!\left\{
j k_0 R
\left[
\sin\theta\,\sin\theta_n
\cos\!\left(\phi-\phi_{(n,k)}\right)
+
\cos\theta\,\cos\theta_n
\right]
\right\}
\end{multline}
Specifically, the component of $e^{-j\Phi_{(n,k)}^{\mathrm{desired}}}$ represents the synthesized transmission phase programmed to steer the radiated beam toward the desired direction, while the second exponential term describes the free-space propagation phase determined by the location of the $(n,k)$-th unit cell on the hemispherical aperture and the observation direction $(\theta,\phi)$. 
Their product ensures coherent superposition of the radiated fields, thereby forming a high-gain beam in the desired steering direction while suppressing radiation in undesired directions.
\begin{figure}[!t]
\centering
\subfloat[]{
\includegraphics[
    width=0.9\columnwidth,
    height=6.9cm,
]{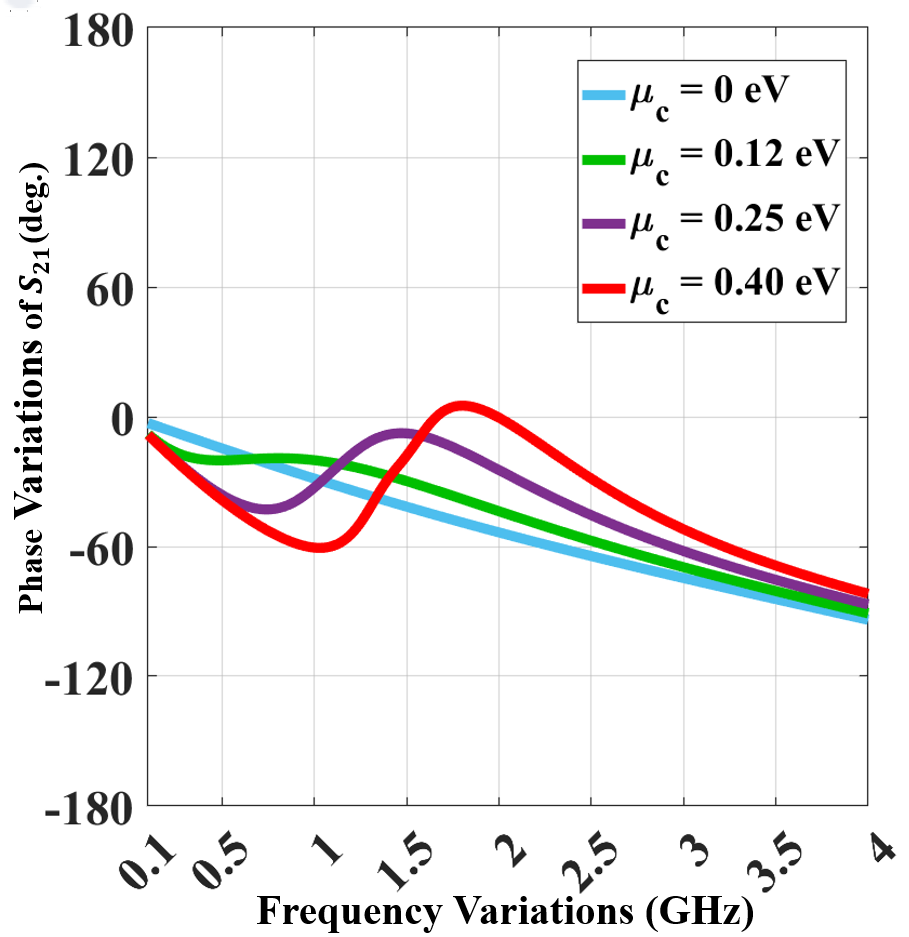}
\label{fig:a2}
}
\vspace{3mm}
\subfloat[]{
\includegraphics[
    width=0.9\columnwidth,
    height=6.9cm,
]{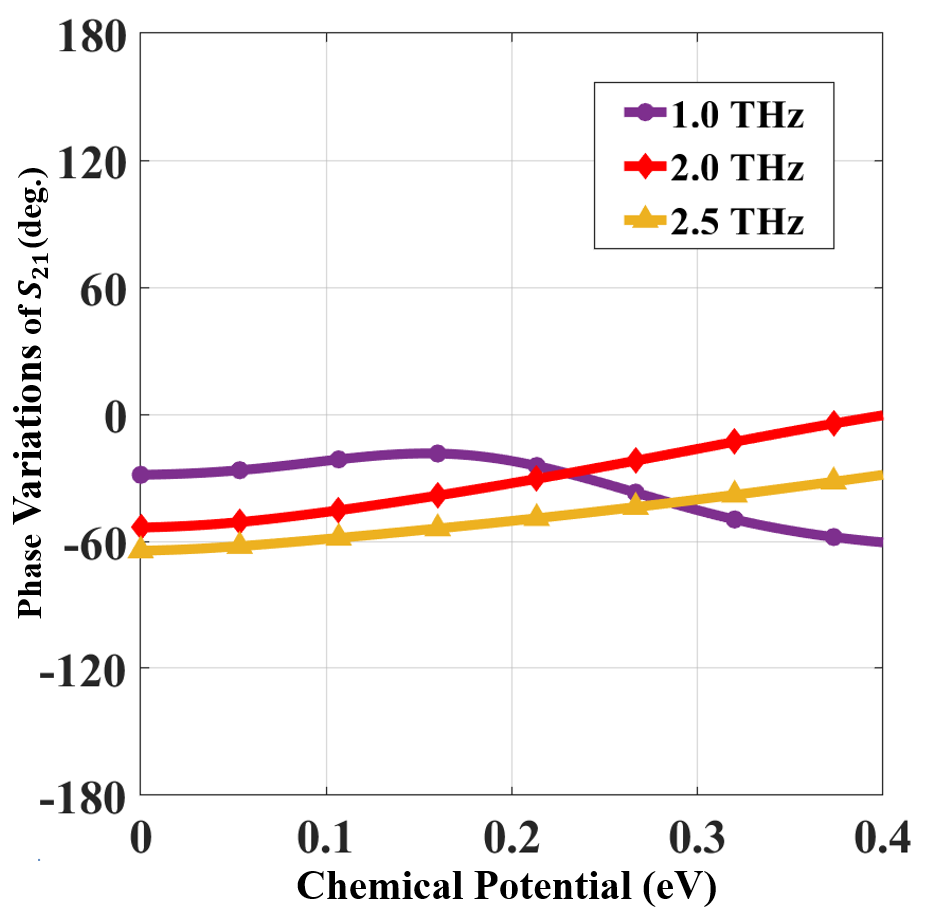}
\label{fig:b2}}
\caption{Phase variations of $S_{21}$ for the proposed hemispherical graphene-based transmitarray antenna: (a) frequency-dependent phase response and (b) chemical-potential-dependent phase response. For each case, a uniform chemical potential is applied simultaneously to all 121 graphene elements to evaluate the overall tunable transmission-phase characteristics of the metasurface.}
\label{fig:S21-Phase Variation}
\end{figure}
\section{Simulation Results and Performance Evaluation}
The proposed graphene-based hemispherical transmitarray antenna in Fig.1 is validated using the full-wave electromagnetic solver CST microwave Studio. 
In the CST Studio Suite environment, the hemispherical transmitarray antenna is excited by a circular feed horn located at the center of the hemispherical lens, providing a realistic illumination with nonuniform amplitude and phase distributions across the metasurface aperture. 
The simulation results obtained from CST microwave Studio were exported to MATLAB for post-processing, analysis, and graphical visualization.
\begin{figure}[!t]
\centering
\subfloat[]{%
\includegraphics[width=0.49\columnwidth]{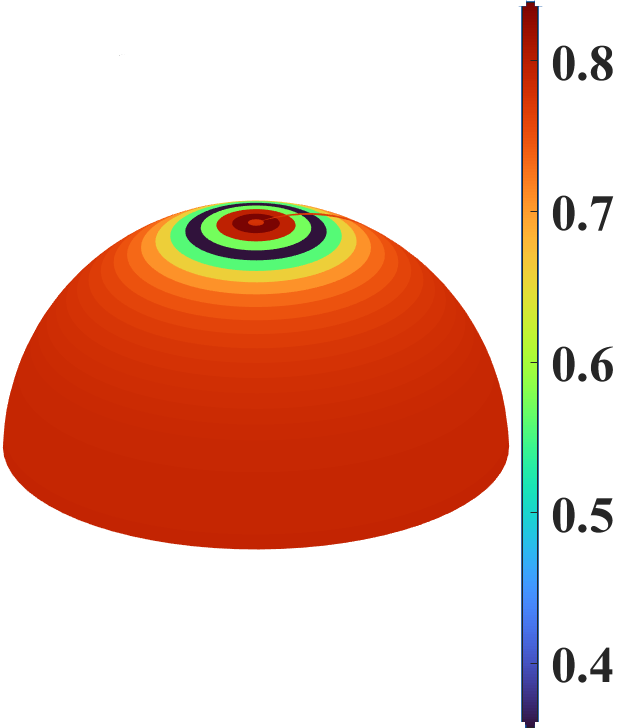}
\label{fig:a}}
\hfill
\subfloat[]{%
\includegraphics[width=0.48\columnwidth]{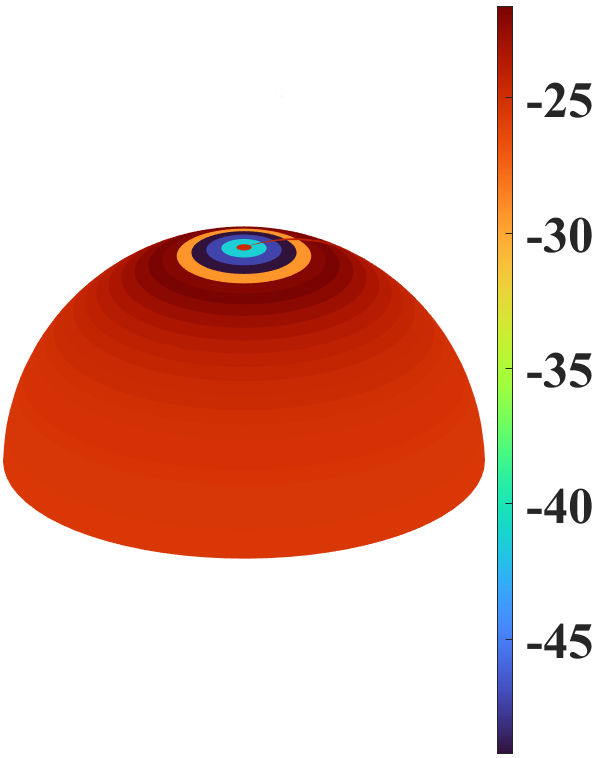}
\label{fig:b}}
\vspace{2mm}
\caption{Distribution of the transmission coefficient of $S_{21}$ over our proposed hemispherical graphene-based transmitarray antenna of Fig.1. In this case, a uniform graphene chemical potential of $\mu_c=0.25$ eV is assigned to all 121 graphene sectors.  The hemispherical transmitarray antenna is illuminated by a circular feedhorn with an aperture size of 5 mm in the CST Studio Suite environment to excite the curved lens. The obtained results $S_{21}$ are subsequently exported to MATLAB for post-processing and visualization of the transmission characteristics across our overall hemispherical transmitarray antenna: (a) transmission-amplitude, $|S_{21}|$, distribution and (b) transmission-phase, $\angle S_{21}$, distribution.}
\label{fig:S21 Distribution}
\end{figure}
The phase variations of $S_{21}$ with respect to frequency and graphene chemical potential are presented in Fig. 8. These results demonstrate the tunable phase response of the graphene-based unit cells and highlight the capability of the proposed transmitarray architecture for dynamic electromagnetic wavefront control.
\begin{figure}[!t]
\centering
\includegraphics[
    width=1.02\columnwidth,
    height=6.2cm,
]{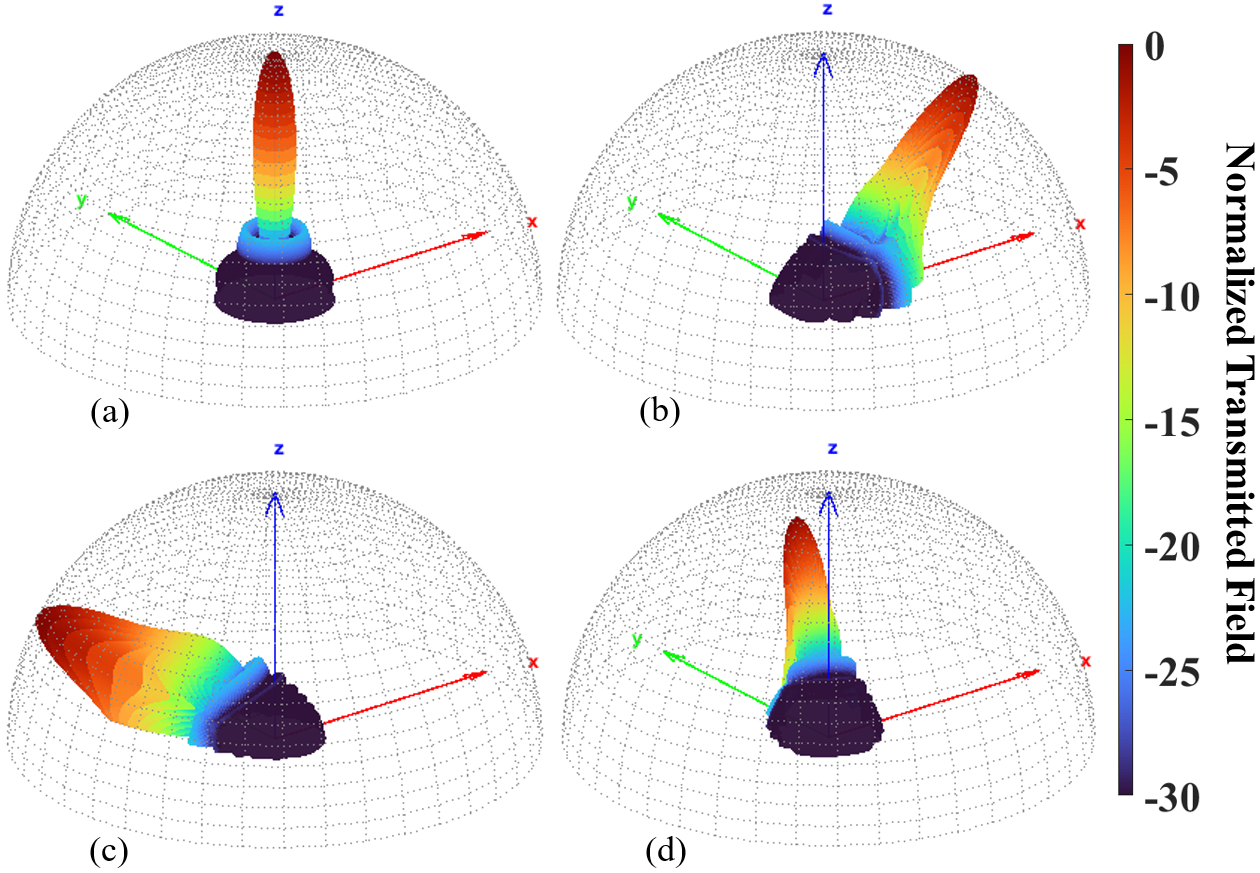}
\caption{Three-dimensional distribution of the normalized transmitted field over the proposed hemispherical transmitarray aperture for evaluating beam-scanning performance along the $\theta$ and $\phi$ angular directions:
(a) $\theta=0^\circ,\phi=0^\circ$, 
(b) $\theta=45^\circ,\phi=0^\circ$, 
(c) $\theta=70^\circ,\phi=120^\circ$, and 
(d) $\theta=60^\circ,\phi=60^\circ$.}
    \label{fig:scanning angles}
\end{figure}
\begin{figure}[!t]
\centering
\includegraphics[
    width=1.02\columnwidth,
    height=6.2cm,
]{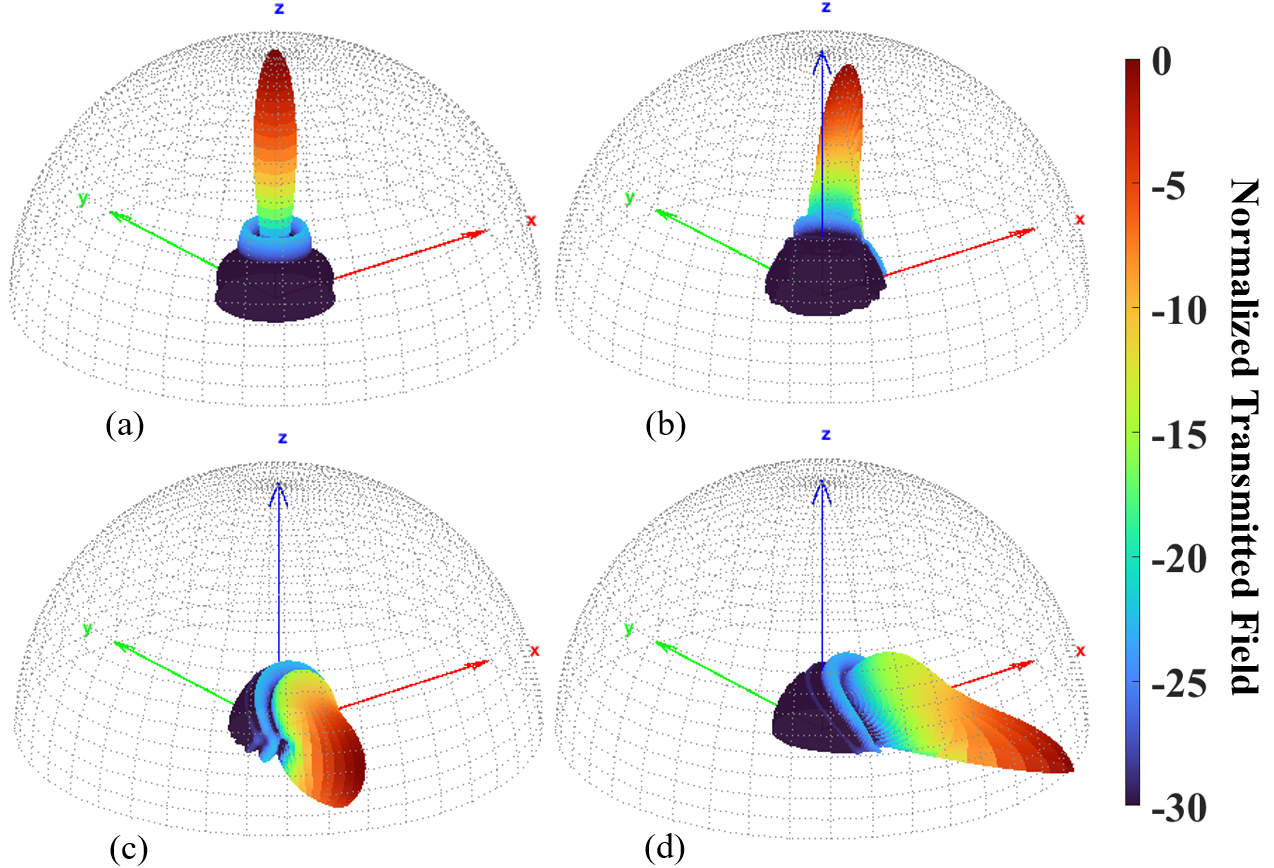}
\caption{Three-dimensional distribution of the normalized transmitted field over the proposed hemispherical transmitarray aperture for evaluating beam-scanning performance along the $\theta$ and $\phi$ angular directions:
(a) $\theta=0^\circ,\phi=0^\circ$, 
(b) $\theta=45^\circ,\phi=45^\circ$, 
(c) $\theta=75^\circ,\phi=250^\circ$, and 
(d) $\theta=90^\circ,\phi=300^\circ$.}
    \label{fig:scanning angles-1}
\end{figure}
Simulation results in Figs. 9 and 10 have verified that the proposed hemispherical transmitarray antenna enables wide-angle beam steering, covering approximately $90^\circ$  in the elevation direction and a full $360^\circ$ range in the azimuth direction. Such comprehensive spatial coverage highlights the advantage of the hemispherical geometry and voltage-controlled graphene reconfiguration over conventional planar metasurface antennas reported in the literature in Table~\ref{tab:beam-reconfiguration-methods}.
\section{Lens Transmission Efficiency}
For the proposed hemispherical transmitarray lens, consistent with Fig.~\ref{fig:transmitarray_antenna}, the local transmission efficiency is governed by the transmission coefficient of $S_{21,(n,k)}$ for the $(n,k)$-th unit cell, which defines the local transmission amplitude and phase across the hemispherical aperture,
\begin{equation}
\eta_{\mathrm{cell},(n,k)}
\approx
\left|S_{21,(n,k)}\right|^2
\end{equation}
This quantity represents the fraction of incident power locally transmitted by the cell, under the assumptions of the equivalent-circuit or full-wave unit-cell model.
It is worth noting that \(\left|S_{21,(n,k)}\right|^2\) quantifies only the local transmission efficiency of the $(n,k)$-th unit cell, without accounting for horn-feed mismatch, aperture spillover, overall radiation efficiency, beam-pointing losses, or sidelobe power distribution. Therefore,
\(\left|S_{21,(n,k)}\right|^2\) should be interpreted as the local transmission efficiency of the $(n,k)$-th transmitarray unit cell.
For our proposed hemispherical transmitarray lens with 121 gold patches, the overall transmission efficiency is expended in terms of the horn-illumination-weighted average of the local transmitted power over the entire aperture given by, 
\begin{equation}
\eta_{\mathrm{lens}}
\approx
\frac{
\displaystyle
\sum_{(n,k)=1}^{m}
\Delta S_{(n,k)}
\left|F_{\mathrm{horn},(n,k)}\right|^2
\left|S_{21,(n,k)}\right|^2
}{
\displaystyle
\sum_{(n,k)=1}^{m}
\Delta S_{(n,k)}
\left|F_{\mathrm{horn},(n,k)}\right|^2
}
\end{equation}
where m denotes the total number of graphene-based unit cells. The complex incident field (or illumination amplitude) of the horn at the $(n,k)$-th unit cell is expressed by $F_{horn,(n,k)}$. The complex transmission coefficient, $S_{21,(n,k)}$, is associated with the corresponding graphene–hBN–gold unit cell. 
The illumination factor of $\Delta S_{(n,k)}\left|F_{\mathrm{horn},(n,k)}\right|^2$ characterizes the nonuniform horn illumination over the hemispherical aperture and represents the power intercepted by the $(n,k)$-th unit cell.

The metric of $\eta_{lens}$ quantifies how efficiently our metasurface lens transmits the electromagnetic power incident from the circular feedhorn through the hemispherical aperture, providing an estimate of the average transmitted power through the lens under horn illumination. Therefore, it is highly effective for evaluating and comparing different graphene bias states, chemical potential distributions, and different geometrical designs.

The CST Simulation results have verified that the overall antenna efficiency, Table~\ref{tab:graphene_antenna_comparison}, of the proposed hemispherical transmit array antenna ranges from 68\% to 80\% for different graphene voltage excitation states. Each bias configuration results in a specific electromagnetic response and corresponding efficiency level. The reported values conservatively include a 10\% efficiency reduction to account for practical imperfections.The proposed metasurface transmitarray antenna demonstrates a substantial efficiency enhancement compared to typical available graphene-based antenna technologies in Table~\ref{tab:graphene_antenna_comparison}.
\section{Conclusion}
To conclude, this work presents the design, analytical modeling, and electromagnetic characterization of a voltage-controlled hemispherical transmitarray antenna operating at 250 GHz, with the objective of enabling wide-range beam steering for real-time moving-target tracking applications.

We have implemented stacked layers of FCC gold patches integrated with tunable graphene sectors, isolated by hBN dielectric layers, to achieve dynamic control of the transmitted electromagnetic wave and subsequent beam steering through spatially localized chemical-potential modulation and surface-impedance tuning by the following procedure, 
\noindent
\[
\begin{aligned}
\mu_c &\rightarrow \sigma_g(\mu_c) \rightarrow Z_s(\mu_c) \rightarrow 
\text{transmission phase of } \phi_t \rightarrow \\
&\text{beam-steering direction of }(\theta,\phi)
\end{aligned}
\]
\begin{table}[t]
\centering
\scriptsize
\caption{Comparison of graphene-based antenna technologies in terms of antenna efficiency.}
\label{tab:graphene_antenna_comparison}
\renewcommand{\arraystretch}{1.1}
\newlength{\RefCol}
\newlength{\FreqCol}
\newlength{\AntCol}
\newlength{\EffCol}
\setlength{\RefCol}{0.08\columnwidth}
\setlength{\FreqCol}{0.12\columnwidth}
\setlength{\AntCol}{0.24\columnwidth}
\setlength{\EffCol}{0.16\columnwidth}
\begin{tabularx}{\columnwidth}{
L{\RefCol}
L{\FreqCol}
L{\AntCol}
L{\EffCol}
Y}
\toprule
\textbf{Refs.} &
\textbf{Frequency Range} &
\textbf{Antenna Type} &
\textbf{Efficiency} &
\textbf{Reconfiguration Mechanism} \\
\midrule

\cite{dash2020behavior}
&
5.8 GHz, 1--2 THz
&
Planar microstrip patch with parasitic graphene
&
MW: 16--66\%; THz: 26--69\%
&
DC-biased graphene conductivity tuning
\\

\cite{vasubabu2022microscaled}
&
0.276--0.711 THz
&
Tree-shaped graphene-based wideband printed MIMO antenna
&
Not reported (ECC $<0.01$, DG $\approx10$ dB)
&
Chemical-potential tuning
\\

\cite{tripathi2019graphene}
&
THz band (2.32 THz bandwidth)
&
Graphene dipole antenna
&
21.5\%
&
DC-biased graphene conductivity tuning
\\

\cite{akbari2016fabrication}
&
889--950 MHz
&
Printed graphene dipole RFID antenna
&
40\%; 2.18 dBi gain
&
No reconfiguration (passive graphene conductor)
\\

This work
&
200--300 GHz
&
Hemispherical graphene transmitarray with FCC patches
&
Achieves values beyond the 68\%--80\% range
&
Voltage-controlled graphene beam steering
\\

\bottomrule
\end{tabularx}
\end{table}
Independent and distinct chemical potentials ranging from 0.1 eV to 1 eV are assigned to the 121 graphene elements to provide flexible spatial phase manipulation and dynamic beam steering. The hemispherical transmit array antenna has exhibited a significant enhancement in its scanning capability, achieving approximately between $-90^\circ$ and $+90^\circ$ elevation coverage and a full azimuthal scanning range of $360^\circ$, and thereby overcoming the angular limitations commonly associated with conventional planar antennas, as discussed in Table~\ref{tab:beam-reconfiguration-methods}.

Our proposed metasurface transmitarray antenna also exhibits an antenna efficiency ranging from 68\% to 80\%, incorporating a conservative 10\% degradation from the ideal theoretical performance to account for fabrication imperfections. A drastic difference in antenna efficiency is observed between available graphene-based antenna technologies, as summarized in Table~\ref{tab:graphene_antenna_comparison}, and the proposed metasurface transmitarray antenna.

Furthermore, the integration of the proposed antenna with an AMD Xilinx Zynq UltraScale+ RFSoC ZCU670 platform validates the feasibility of real-time bias-matrix control and enables ultrafast adaptive beam steering and target tracking, offering a significant advancement over existing beam-steering technologies in Table~\ref{tab:beam-steering-speeds}.
Indeed, the synergistic combination of hemispherical geometry, voltage-controlled graphene reconfigurability, and scalable FPGA-based bias control provides a promising platform for adaptive beam steering and dynamic target tracking in future 6G wireless systems.
\ifCLASSOPTIONcaptionsoff
  \newpage
\fi
\bibliographystyle{IEEEtran}
\bibliography{references}
\end{document}